\documentclass{jfm_modified}
\usepackage{graphicx}
\usepackage{amsmath, amssymb}
\usepackage[normalem]{ulem}
\usepackage{mathtools}

\pdfmapfile{+rsfso.map}
\DeclareSymbolFont{rsfso}{U}{rsfso}{m}{n}
\DeclareSymbolFontAlphabet{\mathscr}{rsfso}

\usepackage{pdflscape}
\usepackage{afterpage}
\usepackage{flafter}
\usepackage{rotating}
\usepackage{fancyhdr}
\fancypagestyle{lscape}{%
\fancyhf{} 
\fancyfoot[LE]{}
\fancyfoot[LO] {}
 
}

\usepackage{bm}
\usepackage{microtype}

\usepackage[pdfauthor={Philippe H. Trinh}]{hyperref}
\usepackage[usenames, dvipsnames]{xcolor}
\hypersetup{colorlinks=true,linkcolor=MidnightBlue,citecolor=MidnightBlue}

\usepackage{natbib}
\usepackage{array}
\usepackage{booktabs}
\usepackage{multirow}

\usepackage{tikz}

\newcommand*{\ep}{\epsilon}

\renewcommand*{\i}{\mathrm{i}}
\newcommand*{\im}{\mathrm{i}}
\newcommand*{\e}{\mathrm{e}}
\newcommand*{\Oh}{\mathcal{O}}
\renewcommand*{\Re}{\operatorname{Re}}
\renewcommand*{\Im}{\operatorname{Im}}
\newcommand*{\Arg}{\operatorname{Arg}}

\renewcommand*{\H}{\mathcal{H}}
\newcommand*{\PP}{\mathcal{P}}
\newcommand*{\QQ}{\mathcal{Q}}

\newcommand*{\ftau}{\tau_\text{free}}
\newcommand*{\ftheta}{\theta_\text{free}}
\newcommand*{\fftheta}{\hat{\theta}_\text{free}}

\newcommand*{\de}{\operatorname{d\!}{}} 
\newcommand{\dd}[2]{\frac{\de#1}{\de#2}}

\newcommand{\pd}[2]{\frac{\partial#1}{\partial#2}}
\newcommand{\ol}[1]{\overline{#1}}

\newcommand{\A}{\mathrm{A}}
\newcommand{\B}{\mathrm{B}}
\newcommand{\C}{\mathrm{C}}
\newcommand{\D}{\mathrm{D}}
\newcommand{\V}{\mathcal{V}}
\newcommand{\Au}{\underline{\A}}
\newcommand{\Bu}{\underline{\B}}
\newcommand{\Cu}{\underline{\C}}
\newcommand{\Du}{\underline{\D}}

\usepackage{mathabx}
\usepackage{esint} 
\def\Xint#1{\mathchoice
   {\XXint\displaystyle\textstyle{#1}}%
   {\XXint\textstyle\scriptstyle{#1}}%
   {\XXint\scriptstyle\scriptscriptstyle{#1}}%
   {\XXint\scriptscriptstyle\scriptscriptstyle{#1}}%
   \!\int}
\def\XXint#1#2#3{{\setbox0=\hbox{$#1{#2#3}{\int}$}
     \vcenter{\hbox{$#2#3$}}\kern-.5\wd0}}

\def\YYint#1#2#3{{\setbox0=\hbox{$#1{#2#3}{\int}$}
     \vcenter{\hbox{\scalebox{1}[-1]{$#2#3$}}}\kern-.5\wd0}}

\def\dashint{\Xint-}

\title[New singularities for Stokes waves]{New singularities for Stokes waves}

\author[S.C. Crew and P.H. Trinh]{P\ls H\ls I\ls L\ls I\ls P\ls P\ls E\ns H.\ns T\ls R\ls I\ls N\ls H}

\affiliation{
Oxford Centre for Industrial and Applied Mathematics, \\ Mathematical Institute, University of Oxford, Oxford OX2 6GG, UK}

\author[S.C. Crew and P.H. Trinh]%
{S\ls A\ls M\ls U\ls E\ls L\ns C.\ns C\ls R\ls E\ls W$^{1}$ 
\and P\ls H\ls I\ls L\ls I\ls P\ls P\ls E\ns H.\ns T\ls R\ls I\ls N\ls H$^{1, 2}$}

\affiliation{%
$^1$ Lincoln College, University of Oxford, Oxford OX1 3DR, UK \\[\affilskip]
$^2$ Oxford Centre for Industrial and Applied Mathematics, Mathematical Institute, \\University of Oxford, Oxford OX2 6GG, UK}

\pubyear{XXXX}
\volume{YYY}
\pagerange{xxx--xxx}
\date{28 April 2016 [Accepted version]}

\begin{document}

\maketitle

\begin{abstract}
In 1880, Stokes famously demonstrated that the singularity that occurs at the crest of the steepest possible water wave in infinite depth must correspond to a corner of $120^\circ$. Here, the complex velocity scales like $f^{1/3}$ where $f$ is the complex potential. Later in 1973, Grant showed that for any wave away from the steepest configuration, the singularity $f = f^*$ moves into the complex plane, and is of order $(f-f^*)^{1/2}$ (\emph{J. Fluid Mech.}, vol.~59, 1973, pp.~257--262). Grant conjectured that as the highest wave is approached, other singularities must coalesce at the crest so as to cancel the square-root behaviour. Despite recent advances, the complete singularity structure of the Stokes wave is still not well understood. In this work, we develop numerical methods for constructing the Riemann surface that represents the extension of the water wave into the complex plane. We show that a countably infinite number of distinct singularities exists on other branches of the solution, and that these singularities coalesce as Stokes' highest wave is approached. 

\end{abstract}

\section{Introduction} \label{sec:intro}

In his \citeyear{stokes_1880_appendix_b} work, Stokes provided a supremely elegant proof that if the crest of an irrotational wave has a sharp edge, then this edge must necessarily make an angle of $120^\circ$. The argument is as follows: for an ideal fluid in deep water with complex potential $f = \phi + \im \psi$ and spatial variable $z = x + \im y$, the constant-pressure condition on the free surface, $y = \eta(x)$, may be taken as
\begin{equation}  \label{berndim}
\frac{1}{2}\left\lvert \dd{f}{z} \right\rvert^2 + g \eta = 0,
\end{equation}
where $g$ is the gravitational parameter, the corner of the wave is assumed to lie at the origin with $z = 0$ and $\phi = 0$, and the free surface is the streamline $\psi = 0$. In the neighbourhood of the corner, if we assume the complex potential takes the form $f \sim A z^n$, then it necessarily follows from \eqref{berndim} that $n = \frac{3}{2}$. Examination of the angular change in $z$ across the crest yields the desired result. 

In fact, Stokes' demonstration was somewhat unexpected, as it corrected an earlier claim by Rankine. As Stokes explains:
\begin{quotation}
\noindent \emph{In a paper published in the Philosophical Magazine, Vol. XXIX (1865), p.25, Rankine gave an investigation which led him to the conclusion that in the steepest possible oscillatory wave of the irrotational kind, the crests become at the vertex infinitely curved in such a manner that a section of the crest by the plane of motion presents two branches of a curve which meet at a right angle.} \citep[p.225]{stokes_1880_appendix_b}
\end{quotation}

\noindent After correcting Rankine's result, Stokes then wondered whether such a wave of greatest height could be obtained in practice:
\begin{quotation}
\noindent \emph{This however leaves untouched the question whether the disturbance can actually be pushed to the extent of yielding crests with sharp edges \ldots After careful consideration I feel satisfied that there is no such earlier limit, but that we may actually approach as near as we please to the form in which the curvature becomes infinite, and the vertex becomes a multiple point where the two branches with which alone we are concerned enclose an angle of $120^\circ$.} \citep[p.227]{stokes_1880_appendix_b}
\end{quotation}
Stokes' conjecture on the existence of such waves would spark a century-long inquiry, until it was positively affirmed by \cite{toland_1978} with the help of many other foundational works (see the review by \cite{toland_1996} for a more comprehensive listing). Equally important, but distinct from studies on the theoretical existence of such waves, is research that attempts to describe the structure of the wave solutions close to the wave of greatest height. We shall focus on this line of inquiry. 

For any wave with amplitude below that of the greatest, the main crest singularity moves off the free surface and lies on the imaginary axis in the complex plane. It was \cite{grant_1973_the_singularity} who first highlighted the complexity of this process, and primarily demonstrated two key difficulties. First, higher-order corrections to Stokes' local result for the steepest wave are extremely non-trivial and required expansions in transcendental powers (\emph{c.f} \citealt[Appendix A]{trinh_2011_do_waveless} for a similar discussion). Second, for wave solutions away from the steepest configuration, the singularities change in power. As Grant writes:
\begin{quotation} \label{grantquote}
\noindent \emph{Now also consider the effect of increasing the amplitude. At maximum amplitude, $z(f)$ has one singularity, of order $\frac{2}{3}$. For any lesser amplitude, it has singularities only of order $\frac{1}{2}$. The only way a continuous approach to greatest amplitude is possible is for $z$ to have several coalescing singularities.} \citep[p.261]{grant_1973_the_singularity}
\end{quotation}
This final statement, regarding the coalescence of square-root singularities, was frustratingly left unexplored. Although Grant understood (particularly aided by the numerical results of \citealt{schwartz_thesis, schwartz_1974}) that the main singularity would approach the free surface, there was no evidence given of any supplemental singularities, nor any explanation of the mechanism by which the local power would change. 

The search for such singularities would not be straightforward. Previously, the location of the main singularity was identified using techniques in series acceleration, such as in the work of \cite{schwartz_1974} and \cite{longuet-higgins_1978_theory_of}. As explained in \citeauthor{baker_book} (\citeyear{baker_book}, p.44) branch points in an analytically continued function will typically manifest as sequences of poles in a Pad\'{e} approximant. Thus, numerical results on the distribution of poles within a series-accelerated approximation can often be deciphered in order to intuit the presence of a branch-point singularity. Recently, \cite{dyachenko_2014, dyachenko_2015branch}, performed extensive numerical computations of the Stokes wave with great attention to the numerical error; these results were then used to construct the Pad\'{e} approximants, and then reconstruct Grant's main singularity.

Such series acceleration techniques can be unreliable or difficult to interpret [see \emph{e.g.} p.20 of \cite{drennan_thesis} and \cite{dallaston_2010_accurate_series}]. Indeed, as it pertains to the Stokes wave, following the early efforts of \cite{schwartz_1974}, \cite{tanveer_1991} had described the situation as follows:
\begin{quotation}
\noindent \emph{However Schwartz's apparent conclusion on the form of the nearest singularity \ldots is in error since it suggested that the branch point power changes continuously between $\frac{1}{2}$ and $\frac{1}{3}$. This erroneous conclusion highlights the difficulty of extracting reliable and accurate information from a Pad\'{e} approximate method [which] is practically feasible only for the nearest singularity and just to leading order.} \citep[p.140]{tanveer_1991}
\end{quotation}

Using an alternative method of analytic continuation, \cite{tanveer_1991} calculated precise estimates of the main singularity on the imaginary axis, and based on the form of the associated free-surface equation, further conjectured the existence of a mirror singularity, which lies on the negative imaginary axis in the adjacent unphysical Riemann sheet. In \S\ref{sec:sing}, we will review Tanveer's reason for believing the existence of the mirror singularity. 

Although Tanveer's work had confirmed the presence of the main singularity and conjectured the existence of its mirrored pair, the work did not comment on the elaborate Riemann sheet structure that would result from the combined influence of these two points. During a recent conference talk, \cite{lush_talk2}, indicated that an infinite number of Riemann sheets would emerge if both singularities were continuously encircled in an alternating manner; these results have now appeared in a preprint \citep{lushnikov2015branch}. 

The main goal of this paper is to construct the Riemann surface that represents the extension of the water wave into the complex plane; in this way, we shall develop a more complete picture of the singularity structure. \emph{Put very simply, we want to see the singularities.} We compare and contrast various numerical methods for achieving this analytic continuation, and demonstrate that in fact, the singularity structure of a finite amplitude wave in deep water is much more complicated than previously anticipated. In particular, not only do we confirm the existence of Tanveer's \citeyearpar{tanveer_1991} mirror singularity, but we also demonstrate the numerical existence of further square-root singularities, shifted in a diagonal fashion, away from the imaginary axis, and lying on subsequent Riemann sheets of the water wave. 

As we explain next, the appearance of such diagonally-shifted singularities is often a generic phenomena associated with nonlinear differential equations (\emph{cf.} \citealt{chapman_2013_exponential_asymptotics}). Moreover, singularities that arise on an unphysical Riemann sheet have been noted in other free-surface flows as well. We highlight, for example, the case of the wave-structure interaction studied in Appendix of \cite{trinh_2013a_new_gravity-capillary}, where such singularities were crucially related to the subsequent analysis of the resultant water waves. 

\subsection{The importance of understanding singularities} \label{sec:singimportance}

The characterization of the singularity structure of the Stokes wave, or indeed any solution of a singular differential equation, brings with it several advantages \citep{tanveer_1991}. First, knowledge of the set of singularities can lead to the development of better analytical methods, and in rare cases, even exact solutions. In the limiting configuration of a steep wave, the curvature near the crest is governed by the effects of the nearest singularities. Thus, the corresponding asymptotic analysis will often require detailed information regarding the type and distribution of such points. In \S\ref{sec:discussion}, we shall discuss the implications of our work with the previous matched asymptotics procedure of \cite{longuet-higgins_1977_theory_of,longuet-higgins_1978_theory_of} for the `almost-highest wave'. 

Second, deeper understanding of the singularity structure can lead to better numerical algorithms for computing waves. Although Stokes' classical approach is sufficient for calculating the profiles of moderate waves [as performed by \emph{e.g.} \cite{schwartz_1974} and \cite{cokelet_1977_steep_gravity}], the presence of the crest singularity severely limits the radius of convergence of Stokes' series representation. In order to compute steep waves, it becomes essential to adopt a different series expansion that accounts for the change in the power of the main crest singularity [\emph{cf.} \cite{olfe_1980_some_new,vanden-broeck_1986_steep_gravity}]. In theory, if we know the detailed locations and scalings of the set of singularities, we can subtract any undesired behaviour from the solution and extend the radius of convergence. 

Finally, and perhaps of widest significance, the importance of knowing the steady-state singularities extends to problems of the time-dependent type as well. As explained by \cite{tanveer_1991}, the occurrence of finite-time interfacial singularities in many free-surface problems is typically associated with singularities in the unphysical complex plane hitting the real domain. For instance, most directly connected with the Stokes wave is the study of how the approach of the nearest crest singularity to the free surface is associated with the breaking of the wave [\emph{cf.} \cite{schwartz_rev_1982, cokelet_1977_steep_gravity, baker_2011_singularities_in}]. However, the role played by complex singularities in the development of physical instabilities is a generic one, and there are a wide range of applications, ranging from Kelvin-Helmholtz and Rayleigh-Taylor instabilities, to rupture and pinch-off phenomena [see \emph{e.g.} \citealt{tanveer_1991} and \cite{chapman_2013_exponential_asymptotics}]. 

\section{Mathematical formulation} \label{sec:mathform}

\noindent Let us consider symmetrical two-dimensional periodic waves of wavelength $\lambda$ under the influence of gravity $g$ at the surface of a fluid of infinite depth. The waves move from right to left with constant speed $c$ relative to an inertial frame, so we shall take a frame of reference moving with the waves and seek steady solutions. The fluid is assumed to be inviscid and incompressible, and the motion irrotational; thus the solution may be represented as an analytic function, $z(f)$, of the complex spatial variable $z = x + \im y$ and complex potential $f = \phi + \im \psi$, where $\phi$ is the velocity potential, and $\psi$ the streamfunction. 

The reader will note the standard reversal of the independent and dependent variables. Although \cite{stokes_1847a_on_the} originally developed his method in the natural way, seeking to solve $f$ as a function of $z$, he published a supplement several decades later \citep{stokes_1880_appendices_and} where he detailed the much simpler inverse formulation. The elegance of the trick relies on the fact that when solving for $z(f)$, the boundary conditions are imposed at a known location in the $f$-plane. Our mathematical presentation follows closely from those formulations discussed in \cite{wehausen_1960}, \cite{schwartz_1974}, \cite{vanden-broeck_1986_steep_gravity}, and \cite{vb_book}.

The Cartesian coordinates are chosen with the $x$-axis at the mean water level, and the $y$-axis is directed vertically upwards against the direction of gravity. The free surface is chosen to be the constant streamline $\psi = 0$, and $\phi = 0$ at the wave crest. The phase velocity, $c$, is defined to be the average horizontal velocity at a constant level, $y$, of the fluid. That is,
\begin{equation}
  c = \frac{1}{\lambda} \int_0^\lambda \pd{\phi}{x} \, \de{x} = \frac{1}{\lambda} \phi(\lambda, y),
\end{equation}
and consequently, it follows from the assumed symmetry and periodicity of the wave that $\phi(\frac{n\lambda}{2}, y) = \frac{cn\lambda}{2}$ for integers $n$. Thus, we shall seek to solve for the fluid within the single periodic domain $-\frac{c\lambda}{2} \leq \phi \leq \frac{c\lambda}{2}$ and $\psi \leq 0$. 

The system is nondimensionalized using $\lambda$ as the unit length and $c$ as the unit velocity; henceforth, all variables are dimensionless. The flow in the physical and potential planes is shown in Fig.~\ref{fig:stokesprofile}(a) and (c). We have shifted the vertical axis of the physical plane for a better visualization of the singularities to be studied in later sections. We write Bernoulli's equation in complex conjugate form (\emph{c.f} \citealt[\S{34}]{wehausen_1960}) as
\begin{equation} \label{eq:bernoulli_mu}
\frac{1}{2} + \frac{2\pi}{\mu} \Im(z) z'(\phi) \ol{z'(\phi)} = B z'(\phi) \ol{z'(\phi)},
\end{equation}
for $-\frac{1}{2} \leq \phi \leq \frac{1}{2}$ and $\psi = 0$, and where the overbar denotes complex conjugation. The introduction of the Bernoulli constant, $B$, differs from \eqref{berndim} due to the chosen mean level, and we have defined
\begin{equation}
\mu = \frac{c^2}{g(\lambda/2\pi)}
\end{equation}
for the square of the Froude number, representing the balance between inertial and gravitational effects. 

\afterpage{
\clearpage
\begin{landscape}
\thispagestyle{lscape}
\pagestyle{lscape}
\begin{figure}
\includegraphics[angle=90]{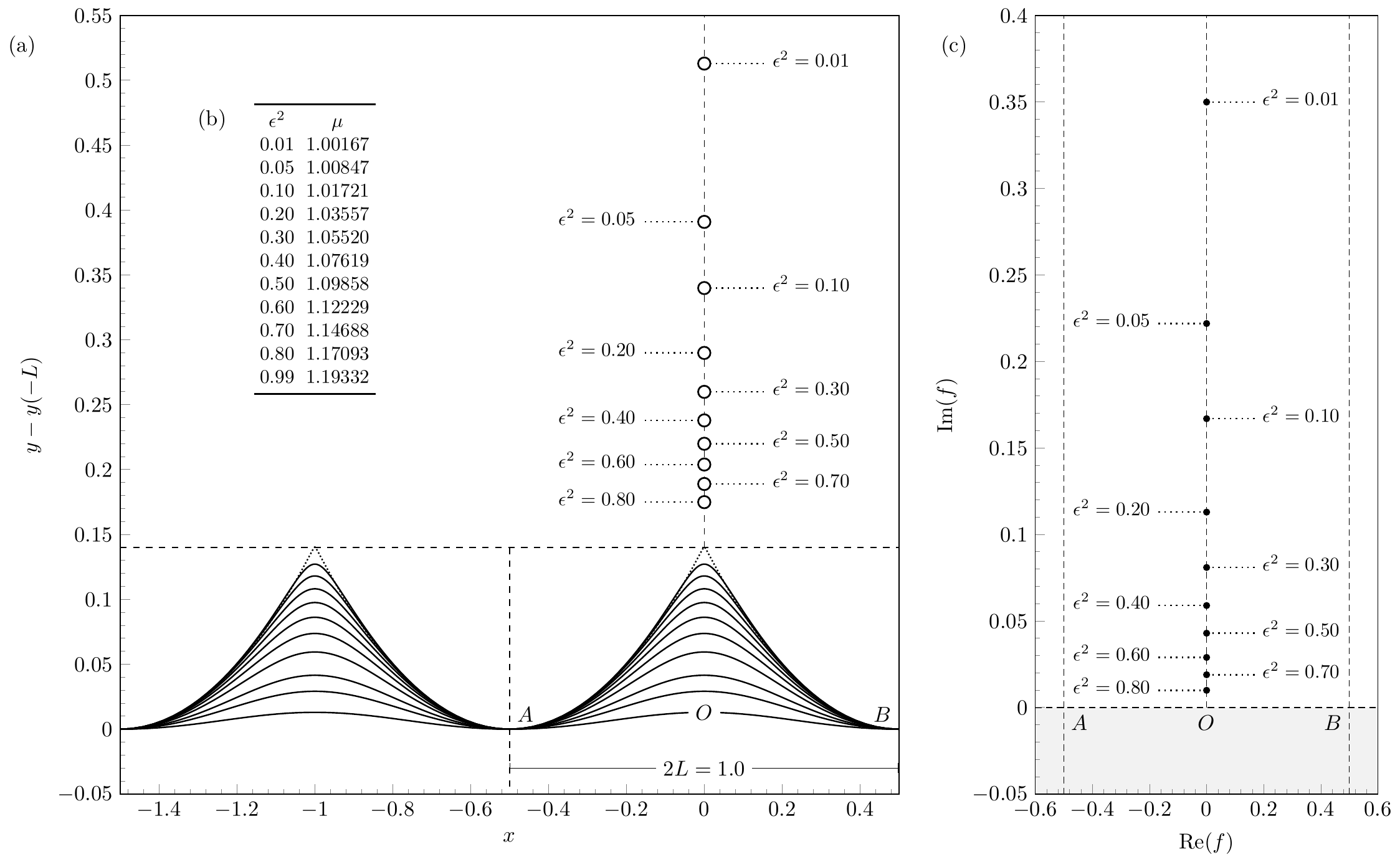}
\caption{(a) Profiles of numerical computed Stokes waves for $\ep$ ranging from 0.01 to 0.99 (increasing in steepness). The $\ep = 0.99$ wave is shown dotted. Notice that rather than centering the mean height of all the profiles, the vertical axis has been shifted. The nodes are singularities in analytically continued plane; (b) values of $\ep$ versus $\mu$; (c) singularities shown in the complex $f$ plane. \label{fig:stokesprofile}}
\end{figure}
\end{landscape}
}

A solution of the nonlinear water wave problem then consists of solving \eqref{eq:bernoulli_mu} for an analytic function $z = z(f)$ in $\psi \leq 0$, such that $z$ is periodic and the velocity is uniform at infinite depth. This requires that $z(f + 1) = z(f)$ and $z'(f) \to 1$ as $\psi \to -\infty$. 

One approach is to posit a series expansion for the solution. For example, following \cite{stokes_1880_appendices_and}, we may set
\begin{equation} \label{zp_series}
  z'(f) = 1 + \sum_{n=1}^\infty a_n \e^{-2 \pi \i n f},
\end{equation}
and then impose the boundary conditions. This determines the coefficients $a_n$, which are assumed to be real by the imposed symmetry of the wave. 

The following numerical scheme is described by \cite{vb_book}. First, the infinite series \eqref{zp_series} is truncated after $N-1$ terms. Next, the $N+1$ unknowns $B$, $\mu$, and $a_n$, $n = 1, \ldots, N-1$, are determined by applying  Bernoulli's equation \eqref{eq:bernoulli_mu} at $N$ equally-spaced points in a single periodic domain. A final equation fixes the amplitude of the wave, and the resultant nonlinear system of $N+1$ equations is solved using \emph{e.g.} Newton's method. 

In regards to the amplitude condition, in this paper, we enforced a fixed value of $\ep$ defined by
\begin{equation}
  \ep^2 = 1 - \left\lvert z'(0) z'(\tfrac{1}{2})\right\rvert^{-2},
\end{equation}
the same convention as in \cite{vanden-broeck_1986_steep_gravity}. In the limit $\ep \to 0$, the wave height decreases to zero, and $\ep = 1$ corresponds to Stokes' greatest wave. As noted by \citeauthor{vanden-broeck_1986_steep_gravity}, the series expansion of \eqref{zp_series} provides an accurate approximation to the wave solutions for $\ep^2$ less than roughly $0.6$. We will mainly limit ourselves to waves that can be accurately calculated using \eqref{zp_series}. 

It was shown by \cite{chen_1980} that gravity waves of infinite depth are not unique, and the classical Stokes wave can bifurcate into new families of solutions at large values of $\ep$. A similar result for finite depth was confirmed by \cite{vanden-broeck_1983_some_new}. These irregular waves are distinguished by having more than one crest per wavelength. In this paper, we focus on the classic Stokes wave (and thus one crest per wavelength). 

Numerical computations of the Stokes waves using \eqref{zp_series} are presented in Fig.~\ref{fig:stokesprofile}(a) for steepness values in the range $0.01 \leq \ep^2 \leq 0.80$. The table of $\ep$ versus $\mu$ values in Fig.~\ref{fig:stokesprofile}(b) confirms that we have achieved at least five-decimal accuracy in comparison with numbers reported in \cite{cokelet_1977_steep_gravity} and \cite{vanden-broeck_1986_steep_gravity}. The dashed curve in Fig.~\ref{fig:stokesprofile}(a) corresponds to the wave with $\ep^2 = 0.99$, which is computed using the generalized Havelock method detailed in \cite{vanden-broeck_1986_steep_gravity}. The crest singularities shown in the figure will be explained in the forthcoming sections.

\section{Preliminary theory of the singularities} \label{sec:sing}

The case of $\ep = 1$ corresponds to Stokes' greatest wave and for this configuration, the singularity is located at the corner $f = 0$ of the $120^\circ$ crest. For any $\ep < 1$, the singularity moves into the upper half-complex-plane, say, at $f = f_{\A}$; we will continue to refer to this point as the \emph{main crest singularity}. Let us first review Grant's \citeyearpar{grant_1973_the_singularity} approach to studying $f_{\A}$, and in particular, the argument that if such a singularity exists, then it must be of square-root type---here, by \emph{square-root type}, we mean that the local power of a solution component (such as $z$ or $z'$) is $\frac{m}{2}$ for some integer $m$. 

\subsection{On the crest singularity and its mirror image} \label{sec:singcrest}

Our first objective is to explain how the free surface, where $f = \phi \in \mathbb{R}$, can be analytically continued into the upper half-plane. By the assumed symmetry of the free surface about the crest, $\phi = 0$, we have the following relation between $z$ and its conjugate:
\begin{equation} \label{eq:zbar}
  \ol{z(\phi)} = -z(-\phi),
\end{equation}
valid for real values of $\phi$.

Next we apply the symmetry condition \eqref{eq:zbar} to Bernoulli's equation \eqref{eq:bernoulli_mu} and reduce the conjugate relations. The resultant equation can then be used as a prescription for continuing into the rest of the complex $\phi$-plane. We write the free-surface potential as $\phi = \phi_r + \i \phi_c$ and, abusing notation somewhat, we relabel $\phi \mapsto f \in \mathbb{C}$. After simplification, \eqref{eq:bernoulli_mu} yields
\begin{equation}  \label{eq:bernoulli_reflected_hidden}
 \QQ(f) \cdot \PP(f) \cdot z'(f) = \mu,    
\end{equation}
where we have introduced the functions
\begin{equation} \label{PQ}
\PP(f) \equiv \mu B + \im \pi [z(-f) + z(f)] \quad \text{and} \quad \QQ(f) \equiv 2z'(-f).
\end{equation}

When $f$ is in the upper half-plane, then those components in $\PP$ and $\QQ$ that require values of $z$ and $z'$ at $-f$ are known from the convergent in-fluid series \eqref{zp_series}. Thus \eqref{eq:bernoulli_reflected_hidden} forms a well-defined initial-value problem for $z$ in the upper half-plane. In \S\ref{sec:analcont}, we shall discuss how \eqref{eq:bernoulli_reflected_hidden} can be used to explore the topology of the Riemann surface of $z$. 

The following dominant balance argument can be verified \emph{a posteriori}. Let us suppose that the singularity lies at $f_{\A}$ in the upper half-plane. Based on the imposed symmetry of the fluid and the form of \eqref{eq:bernoulli_reflected_hidden}, we would then expect $f_{\A}$ to be purely imaginary. Near this point, we assume that the solution can be represented in the form of
\begin{equation} \label{zposit}
  z \sim a_0 + a_1 (f - f_{\A})^\alpha,
\end{equation}
for constants $a_0$ and $a_1$, and $0 < \alpha < 1$. The restriction on $\alpha$ ensures that $z'$ is sufficiently singular. Since $\QQ$ is an analytic function in the upper half-$f$-plane, then at leading-order in \eqref{eq:bernoulli_reflected_hidden}, we have 
\begin{equation} \label{Plimit}
  \PP(f_{\A}) = \mu B + \pi \im [z(-f_{\A}) + z(f_{\A})] = 0,
\end{equation}
which yields a value for $a_0$ in \eqref{zposit}. At next order, we may then use the fact that $\QQ(f_0)$ is bounded and non-zero to conclude that
\begin{equation} \label{alpha12}
   (f - f_\A)^{2\alpha - 1} = \Oh(1) \quad \Rightarrow \quad \alpha = 1/2,
 \end{equation} 
thus establishing the square-root behaviour as remarked by \cite{grant_1973_the_singularity}. 


In fact, it is a consequence of the periodicity of the fluid that there will be further copies of the crest singularity, situated at $f_{\A} + \mathbb{Z}$, with each copy separated by the unit length of the periodic interval. The general properties of the singularities that arise due to the underlying periodicity of \eqref{zp_series} are non-trivial, and we explain these details in \S\ref{sec:global}. For the moment, let us focus on those singularities that are generated within the main interval.

Like Tanveer (\citeyear[p.149]{tanveer_1991}), we may develop the following argument to propose that a mirror singularity might be expected at $-f_{\A}$. Suppose that using \eqref{eq:bernoulli_reflected_hidden}, the solution is analytically continued around the branch cut from $f_{\A}$, and re-enters the lower half-plane; this part of the plane corresponds to an unphysical Riemann sheet, rather than the in-fluid domain. Since $\QQ$ is singular at $-f_{\A}$, then based on the condition \eqref{eq:bernoulli_reflected_hidden}, it is likely that $z$ will also be singular. Although Tanveer did not elaborate beyond this point, we may go further and apply the same dominant balance arguments as above. It may be verified that, unlike the main crest singularity associated with property \eqref{Plimit}, $\PP$ tends to a non-zero constant as $f$ tends to $-f_{\A}$. Using \eqref{alpha12} for $\QQ$, an asymptotic balance for \eqref{eq:bernoulli_reflected_hidden} yields
\begin{equation} \label{alpha32}
  z' = \Oh(f + f_{\A})^{1/2} \quad \Rightarrow \quad
  z \sim b_0 + b_1 (f + f_{\A})^{3/2},
\end{equation}
for constants $b_0$ and $b_1$. The weaker nature of the singularity at the mirror point does not appear to have been remarked upon by others. 

\subsection{On general singularities} \label{sec:singgen}

Later in \S\ref{sec:riemann}, not only do we confirm the existence of the main crest singularity and its mirror image, but in \S\ref{sec:diag}, we will also demonstrate the existence of further singularities: a countably infinite set, diagonally shifted from the positions of $f_{\A}$ and $-f_{\A}$. In fact, the dominant-balance arguments that we have applied to conclude \eqref{alpha12} and \eqref{alpha32} can be repeated for a general singularity, say, at $f = f^*$. 

We propose that there are two types of singularities, whose classification depends on the behaviour of $\PP$ as $f$ approaches $f^*$. The two types are described by
\begin{equation} \label{Psing}
  \PP(f) \to 
  \begin{cases}
  0 & \text{with $z \sim a_0 + a_1 (f - f^*)^{1/2}$}\,, \\
  \text{const.} \neq 0 & \text{with $z \sim b_0 + b_1 (f - f^*)^{3/2}$}\,.
  \end{cases}
\end{equation}

Recall the argument applied to justify the existence of the mirror singularity; that is, a singularity at $f_{\A}$ must produce a non-analyticity in $\PP$ or $\QQ$ at the reflected point $-f_{\A}$. This argument can be similarly applied to the case of a general singularity. Suppose for instance that $f^*$ is a singularity of $\tfrac{1}{2}$ type in \eqref{Psing}, so with $\PP \to 0$. Studying the functional form of $\PP$ in \eqref{PQ}, it would seem unlikely for it, too, also to tend to zero as $f$ tends to $-f^*$ after crossing the lone branch cut from $f^*$; after all, there is no reason for the analytic continuation to be symmetric in this fashion. Therefore, it is likely that the mirror singularity $-f^*$ switches to $\frac{3}{2}$ type. Based on this very informal argument, we intuit that the \emph{chance} existence of a singularity due to $\PP = 0$ can cause the emergence of a further mirror singularity where $\PP \neq 0$. 


We will apply a variety of techniques to compute the locations of the singularities. The most crude approach is simply to observe the values of $z$ and $z'$ along particular curves, surface meshes, or contour plots. Branch cuts appear as differences in levels or kinks within such plots. Alternatively, we can observe numerical unboundedness as the singularity in $z'$ or $1/z'$ is approached. Generally, very approximate locations of the singularities are not difficult to obtain.

However, a more elegant approach can be designed based on contour integration. Near a general singularity, we have, after inversion,
\begin{equation} \label{f0calcstop}
z'(f) \sim d(f-f^*)^\alpha \quad \Rightarrow \quad
f(z') \sim f^* + (z'/d)^{1/\alpha}.
\end{equation}
for constant $d$. We divide the second relation by $z'$ and integrate the result along a closed counterclockwise contour, $C'$, containing $z' = 0$. The second term on the right hand-side of \eqref{f0calcstop} integrates to zero, and this yields
\begin{equation} \label{f0calc}
f^* = \frac{1}{2\pi \i} \oint_{C'} \frac{f}{z'} \, \de{z'} = \pm \frac{1}{2\pi \i}\oint_C f\frac{z''}{z'} \, \de{f}.
\end{equation}
The last equality in \eqref{f0calc} follows from the change of variables with $\de{z'} = z'' \de{f}$, and subsequent integration of the map of the contour, $C$, in the $f$-plane. The $\pm$ sign follows from the ambiguity of not knowing whether a positive orientation in the $z'$-plane is associated with a positive orientation in the $f$-plane. However, once this orientation has been established (\emph{e.g.} for the main crest singularity), the conformal property of complex functions assures us that the sign will not change for other singularities. In our numerical computations to follow, we integrate \eqref{f0calc} using the trapezoid rule, where the values of $f$ and $z'$ are found from the scheme of \S\ref{sec:analcont}, and $z''$ is calculated using centred differences.

Finally, note that based on the dominant balance arguments above, only singularities of square-root type are admissible in the analytic continuation of the Stokes wave. Locally, the Riemann surface near a square-root branch point contains only two sheets (or two branches), and thus positive or negative rotations are equivalent. However, due to the uniqueness of the analytic continuation process, this local property must also be true for the global surface. For instance, consider a point $f$ on the Riemann surface for $z(f)$ which is reached through a positive (counterclockwise) rotation about a given branch point. If we were to replace the positive rotation by the equivalent negative (clockwise) rotation, this process would yield the same value for $z(f)$.

\section{Analytic continuation of the Stokes wave} \label{sec:analcont}

Once the coefficients of the series representation have been computed using the free surface condition \eqref{eq:bernoulli_mu}, the in-fluid values, with $\Im(f) = \psi < 0$ are obtained directly from \eqref{zp_series}. However, the analytic continuation of the free surface to the upper half $f$-plane is severely limited by the main singularity identifiable with the crest of the steepest wave. In the past, authors had attempted to use techniques in series acceleration (primarily Pad\'{e} approximants) in order to extend the radius of convergence of \eqref{zp_series}. However, such techniques are can be unreliable and produce spurious singularities. We now implement two methods for constructing the analytic continuation of the Stokes wave.

\subsection{Method 1: Reflection about the origin} \label{sec:reflect}

The first method of analytic continuation, which we introduced in \S\ref{sec:singcrest}, was originally proposed by \cite{grant_1973_the_singularity} and then extended by \cite{tanveer_1991}. Here, we improve the method to allow for exploration onto further branches of the solution. From \eqref{eq:bernoulli_reflected_hidden}, we have the equation
\begin{equation} \label{eq:bernoulli_reflected}
z'(f) = \frac{1}{2z'(-f)\left\{B - \frac{\pi}{\mu\im} [ z(f) + z(-f)]\right\}}.
\end{equation}

\begin{figure} \centering
\includegraphics{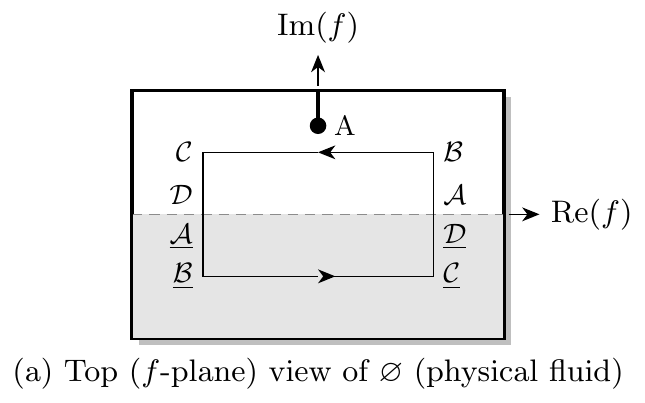}

\includegraphics{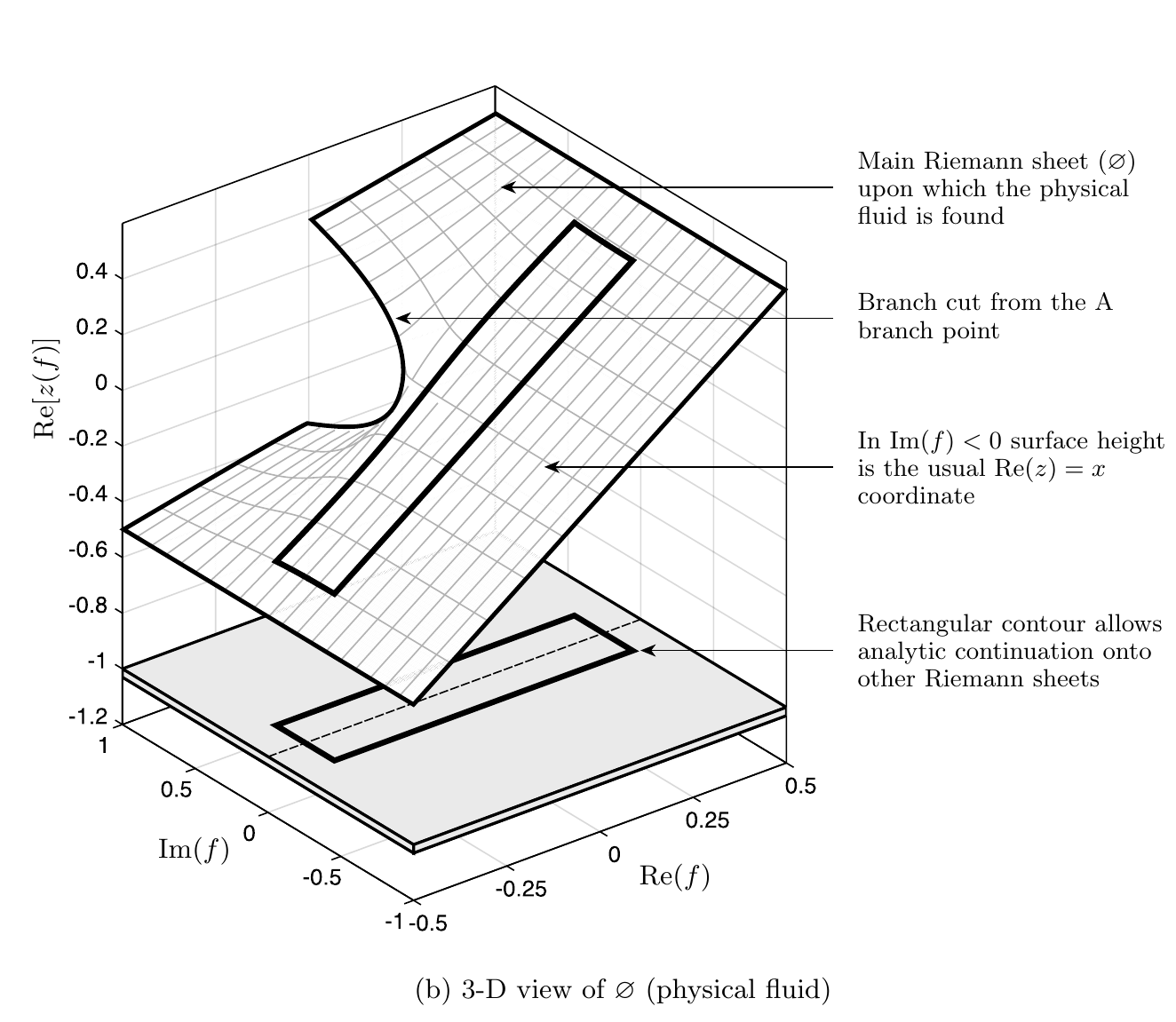}
\caption{Projection of the Riemann surface into $(\Re f, \Im f, \Re z)$-space for the solution with $\ep = 0.1$. The surface is meshed by numerically integrating \eqref{eq:bernoulli_reflected} along vertical paths beginning from the free-surface solution along the $\Re f$ axis. \label{fig:reflection}}
\end{figure}

The simplest way to solve \eqref{eq:bernoulli_reflected} is to begin on the known free surface calculated from \eqref{zp_series}, and integrate upwards along a vertical path in the $f$-plane. For example, consider the path labeled $\mathcal{A}\mathcal{B}$ in Fig.~\ref{fig:reflection}(a). The path is reflected about the origin to $\uline{\mathcal{A}\mathcal{B}}$. Since the values of $z(-f)$ and $z'(-f)$ are known from the series expansion \eqref{zp_series}, then \eqref{eq:bernoulli_reflected} provides an initial-value-problem that can be solved using any standard numerical stepping scheme (in our case, an adaptive Runge-Kutta solver).

We use this method to mesh the $\ep = 0.1$ surface shown in Fig.~\ref{fig:reflection}(b). In the region $\Im(f) < 0$, this yields the physical fluid, which can be verified to match the convergent series values. In the upper half-plane, we observe a clear branch cut and branch point, labeled A, which is the main singularity as conjectured by \cite{grant_1973_the_singularity}. We shall say more about the specifics of the surface and singularity in \S\ref{sec:riemann}.

Now suppose we wish to obtain values on other branches of the solution. Consider the rectangular path labeled $\mathcal{A}\mathcal{B}\mathcal{C}\mathcal{D}$ in Fig.~\ref{fig:reflection}(a). The values of the solution along this rectangular path only depend on the values along $\uline{\mathcal{A}\mathcal{B}\mathcal{C}\mathcal{D}}$, and can thus be determined in an analogous manner, by solving \eqref{eq:bernoulli_reflected} beginning from $\mathcal{A}$. 

Once the values in the upper half-plane are known, the path of continuation can proceed into the lower half-plane as well. If the prior journey had not circled a branch point, then we remain on the same Riemann sheet, and the lower half-plane will be the same as the in-fluid region calculated from \eqref{zp_series}. Indeed, this is the case in Fig.~\ref{fig:reflection}(b) where we observe the rectangular contour lying on the same surface mesh as determined through vertical integration. On the other hand, if a critical point was encircled, the continuation has taken us onto an unphysical Riemann sheet lying `adjacent' to the physical fluid. This continuation process can then be iterated with the values in each lower or upper half-plane reflected in order to provide the values for the next half-plane.

The strength of the reflection scheme for continuation is that it is accurate; for a good approximation to the initial values on the free surface, the only error in the scheme occurs from the integration of \eqref{eq:bernoulli_reflected}, which can be controlled. However, the approach is limited by the requirement that the chosen path for continuation must be symmetric about the origin, and thus it does not allow us to derive the values $z(f)$ for a general trajectory. In our numerical results for \S\ref{sec:riemann}, we implement the reflection method for rectangular contours centred about the origin. 

\subsection{Method 2: Boundary integral continuation} \label{sec:bdint}

We propose an alternative approach that uses the boundary integral formulation of the Stokes wave problem and is not limited by symmetry. Boundary integral schemes provide an alternative treatment of potential free-surface problems, particularly in cases where the problem geometry (here fluid of infinite depth) is more complicated. An extensive review of such techniques can be found in the book by \cite{vb_book}. 

To begin, the complex velocity is written as $\dd{f}{z} = \exp(\tau - \i\theta)$ for the fluid speed $\e^\tau$ and streamline angle by $\theta$. The logarithmic hodograph variable, $\Omega(f)$, is then introduced by
\begin{equation} \label{hodograph}
  \Omega(f) \equiv \tau - \i\theta = \log[1/z'(f)].
\end{equation}

Since $\Omega(f)$ is an analytic function in the fluid region, Cauchy's integral formula applied to a rectangular contour bordering $\phi \in [-0.5, 0.5]$ and $\psi \in (-\infty, 0]$ provides a boundary integral relationship between the real and complex parts of $\Omega$. For values along the free surface, $f = \phi + \im 0$, the boundary integral is given by
\begin{equation} \label{bdint_real}
  \tau(\phi + \im 0) = \dashint_{-0.5}^{0.5} \ftheta(\varphi) \cot\pi(\varphi - \phi) \, \de{\varphi},
\end{equation}
where the integral is of principal value. 

At this point, the integral \eqref{bdint_real}, combined with Bernoulli's equation \eqref{eq:bernoulli_mu}, provides a closed system for the determination of $\tau$ and $\theta$ along the axis, and indeed this forms the basis for such numerical boundary integral schemes. Instead, we will be interested in analytically continuing the integral \eqref{bdint_real} off the axis. The analytic continuation of the governing system gives two equations for $\tau = \tau(f)$ and $\theta = \theta(f)$, valid now for $f\in\mathbb{C}$,
\refstepcounter{equation}\label{analgov}
\[
\e^{3\tau} \dd{\tau}{f} + \frac{2\pi}{\mu} \sin\theta = 0
\quad \text{and} \quad
\tau + a\i \theta = \H(f), \tag{\theequation a,b}
\]
with $a = \pm 1$ for $\Im(f) \gtrless 0$, and where we have defined $\H$ according to
\begin{equation}
  \H(f) \equiv \int_{-0.5}^{0.5} \ftheta(\varphi) \cot\pi(\varphi - f) \, \de{\varphi}.
\end{equation}

The first equation in \eqref{analgov} is the differentiated and analytically continued form of Bernoulli's equation (\emph{cf.} eqn (6) in \citealt{vanden-broeck_1986_steep_gravity}), written in terms of $\tau$ and $\theta$. The second equation is the analytically continued boundary integral, where $a = 1$ for continuation into the upper half-plane and $a = -1$ for the lower half-plane. Notice that upon taking $\Im(f) \to 0$ in the equation, we obtain a principal-value integral and residue contribution, and subsequently return to \eqref{bdint_real}, which indeed verifies that the continuation is done properly. 

Consider a continuation into the upper half-plane that returns down to the real axis. Upon crossing into the lower half-plane, we have
\begin{equation} \label{eq:bdint_rule}
  \tau + \i \theta = \H(f) + 2\i \ftheta(f),
\end{equation}
where $\ftheta$ is the analytic continuation of the free surface angle into the lower half-plane. Now if $\theta$ on the left hand-side of \eqref{eq:bdint_rule} is still on the same Riemann sheet as $\ftheta$, that is to say, a branch point or pole was not encircled during the prior continuation, then $\theta(f) = \ftheta(f)$, and we recover (\ref{analgov}b) with $a = -1$. This would yield the values within the physical fluid region. However, if our analytic continuation took us onto a different Riemann sheet, then \eqref{eq:bdint_rule}, with additional work to find the value of $\ftheta$, is required in order to obtain the values in the lower half-plane.

The analogous rule applies to the case where we wish to continue from \eqref{eq:bdint_rule} back into the upper half-plane. For this, we must now include an additional residue,
\begin{equation} \label{eq:bdint_rule2}
  \tau + \i \theta = \H(f) + 2\i \ftheta(f) - 2\i \fftheta(f).
\end{equation}
If it was the case that no critical point was encircled while in the lower half-plane, we would remain on the same Riemann sheet as in \eqref{eq:bdint_rule}, $\ftheta = \fftheta$, and we thus recover (\ref{analgov}b) with $a = 1$. However, if we have traversed onto the another Riemann sheet, then in general $\ftheta$ will possess different values from $\fftheta$, the latter of which is the analytic continuation of the free surface directly into the adjacent upper half-plane. 

\begin{figure} \centering
\includegraphics{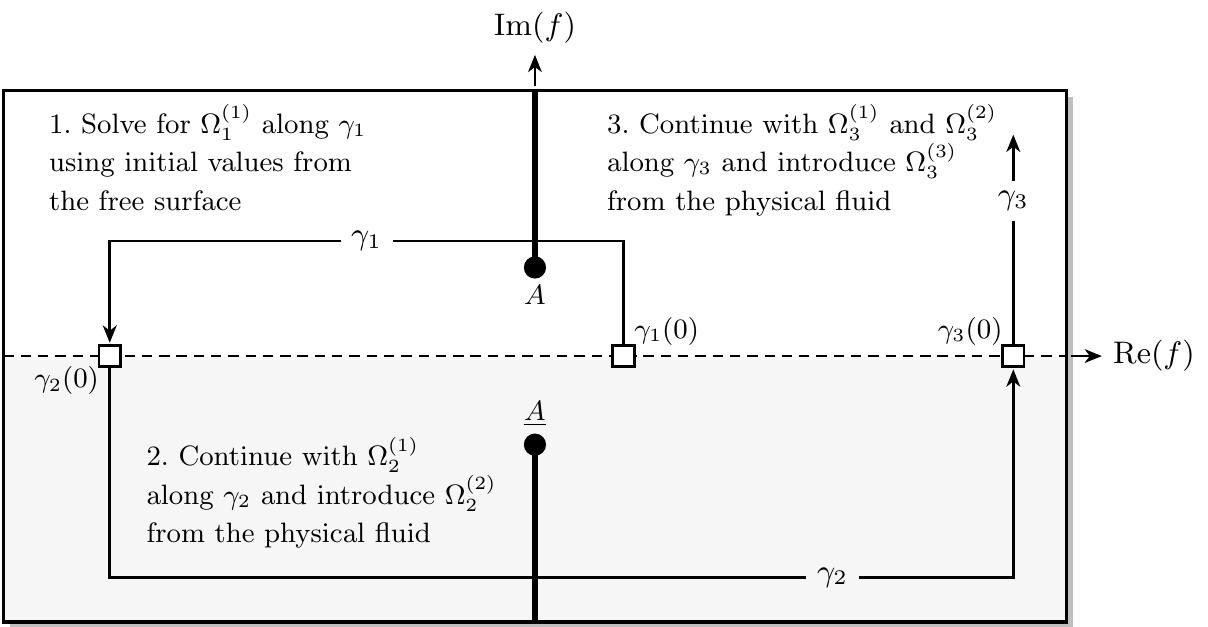}
\caption{Projection of the Riemann surface into $(\Re f, \Im f, \Re \Omega)$-space for the solution with $\ep = 0.1$. The surface is meshed by using the boundary integral continuation. \label{fig:bdcont_contour}}
\end{figure}

Let us now describe the numerical procedure for continuation along an arbitrary path in the $f$-plane, written as $\gamma$, and shown in Fig.~\ref{fig:bdcont_contour}. We first break $\gamma$ into segments each time the axis is crossed, and write 
\begin{equation}
  \gamma = \gamma_1(s) \cup \gamma_2(s) \cup \ldots \gamma_n(s),
  \end{equation}
with each sub-segment parameterized by $s\in[0, 1]$. The first contour, $\gamma_1$ is in the upper half-plane, and the initial point, $\gamma_1(0)$ lies on the free surface. By our notation, $\gamma_i$ lies in the upper half-plane if $i$ is odd and the lower half-plane if $i$ is even. Let $\tau_{i}$, $\theta_{i}$, and $\Omega_i$, be the analytically continued values along the sub-segments, $\gamma_i$. In addition to the subscript notation, we shall use an additional superscript, such as $\Omega_i^{(j)}$, for individualized Riemann sheets, to be explained below.

First, we solve for $\Omega_{1} = \Omega_{1}^{(1)}$ in the upper half-plane using (\ref{analgov}b). Once $\gamma_1$ reaches its end along the real axis, we continue onto $\gamma_2$ in the lower half-plane, introduce the residue contribution \eqref{eq:bdint_rule}. We call this newest contribution, $\Omega_{2}^{(2)}$, that is to say, from \eqref{hodograph} it is composed of $\tau_{2}^{(2)} - \im\theta_{2}^{(2)}$. The quantities $\tau_{2}^{(2)}$ and $\theta_{2}^{(2)}$ correspond to physical fluid quantities, and can be calculated by returning to (\ref{analgov}b) and solving with $a = -1$; more simply, it consists of the in-fluid values directly obtained from the series representation \eqref{zp_series}. 

Once we arrive at the end of $\gamma_2$ and continue onto $\gamma_3$, then again, we must include a separate residue contribution and this involves the introduction of $\Omega_{3}^{(3)}$. Thus, the desired values of $\Omega$ on each section of the plane will depend, in a recursive manner, on the values of the preceeding Riemann sheets. Consequently,
\begin{equation}
  \text{$\Omega_{n}^{(j)}$ depends on $\Omega_{n}^{(n)}$, $\Omega_{n}^{(n-1)}$, \ldots $\Omega_{n}^{(j+1)}$}.
\end{equation}

Then the analytic continuation, $\tau_{n}^{(j)}$ and $\theta_{n}^{(j)}$, along the curve $\gamma_i$ will involve solving the $n^\text{th}$-order system of equations,
\begin{subequations} \label{eq:bdintscheme}
\begin{equation}
 \dd{\tau_{n}^{(j)}}{f} = - \left[\frac{2\pi}{\mu} \sin\theta_i^{(j)}\right] \e^{-3\tau_{n}^{(j)}} \quad \text{ for $1 \leq j \leq n$},
\end{equation}
with the initial conditions,
\begin{align}
\tau_{n}^{(j)}(f) \Bigr\rvert_{f = \gamma_n(0)} &= \ftau(\phi) \Bigr\rvert_{\phi = \gamma_n(0)}
 && \text{for $j = n$}, \label{taubc1} \\
\tau_{n}^{(j)}(f)\Bigr\rvert_{f = \gamma_n(0)} &= \tau_{(n-1)}^{(j)}(f)\Bigr\rvert_{f = \gamma_{n-1}(1)} && \text{for $1 \leq j \leq n-1$}. \label{taubc2}
\end{align}
The initial condition in \eqref{taubc1} imposes that the newest residue contribution begins from the physical free surface, $\ftau$, calculated from \eqref{zp_series} and \eqref{hodograph}. The initial conditions \eqref{taubc2} impose that the previously generated contributions all continue from the end of the preceding path.

The system of differential equations is then closed with the system of recursively defined algebraic equations for $\theta$ given by
\begin{equation}
 (-1)^{n+1} \i \theta_{n}^{(n)} = \H(f) - \tau_{n}^{(n)}, \qquad \text{for $n \geq 1$,} 
\end{equation}
and
\begin{equation} \label{thetanj}
(-1)^{j+1} \i \theta_{n}^{(j)} = \H(f) - \tau_{n}^{(j)} + 2\i \sum_{k = j + 1}^n (-1)^k \theta_{n}^{(k)},
 \quad \text{for $1 \leq j \leq n-1$ and $n \geq 2$.}    
\end{equation}
\end{subequations}

Given a path, $\gamma = \gamma_1 \cup \ldots \cup \gamma_n$, the system \eqref{eq:bdintscheme} can be solved using any standard ordinary differential equations scheme (in our case, an adaptive Runge-Kutta solver). 

\begin{figure} \centering
\includegraphics{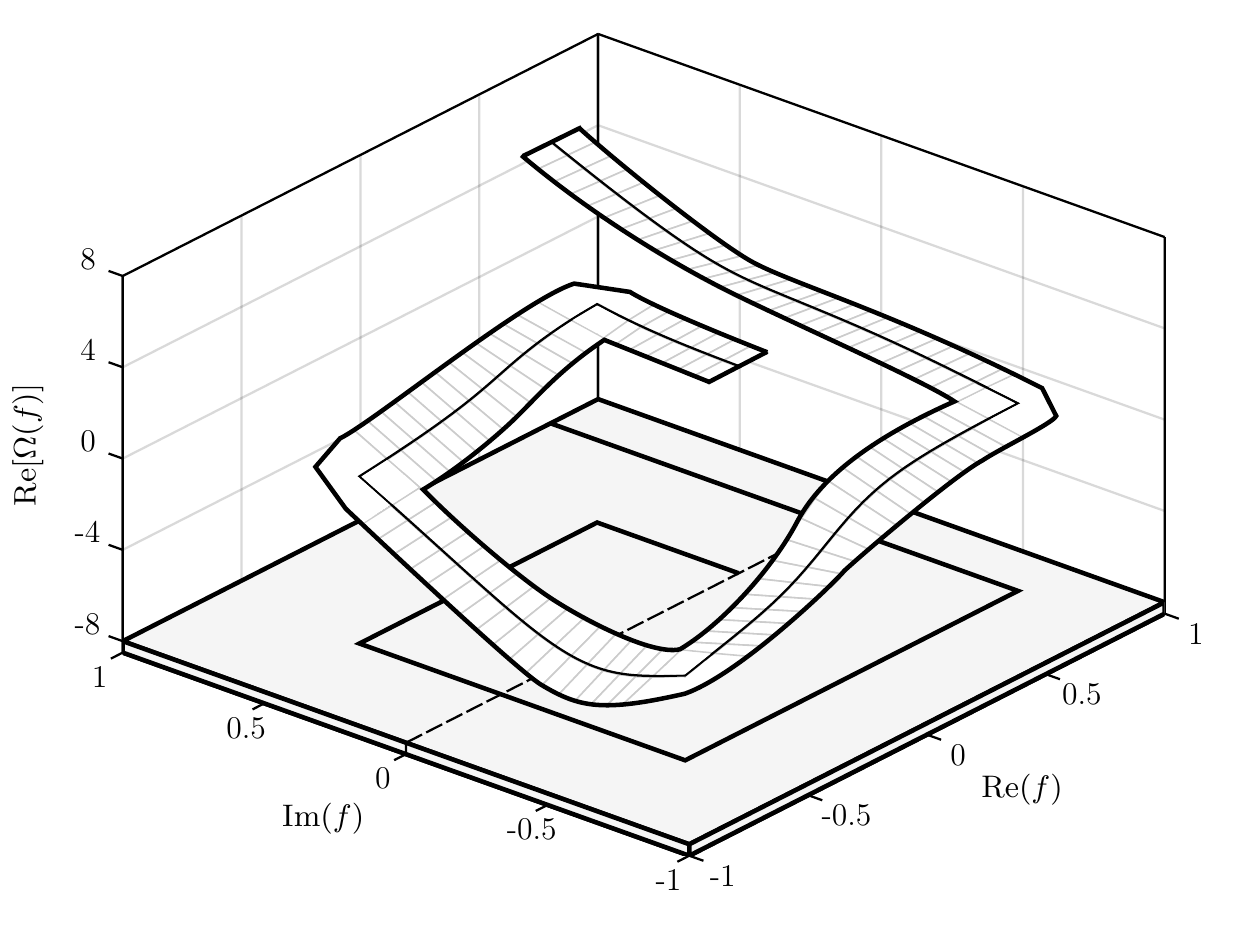}
\caption{Projection of the Riemann surface into $(\Re f, \Im f, \Re \Omega)$-space for the solution with $\ep = 0.1$. The surface is meshed by using the boundary integral continuation. \label{fig:bdcont_surface}}
\end{figure}

The result of using the boundary-integral continuation with the path shown in Fig.~\ref{fig:bdcont_contour} is displayed in Fig.~\ref{fig:bdcont_surface} for the Stokes wave with $\ep = 0.1$. The center-line of the surface mesh in Fig.~\ref{fig:bdcont_surface} is the exact computed value, and for visualization purposes, we have extended the center-line into a surface mesh in the following way: at each point, we construct a planar patch using a vector, $\bm{v}_1$, along the arc-length direction, and its normal, $\bm{v}_2$, assumed to satisfy conformality of the function $\Omega(f)$. Unscaled, these are written
\begin{equation}
 \bm{v}_1 = \Bigl[\Re (\gamma'), \Im (\gamma'), \Re (\Omega_s)\Bigr] \quad 
 \text{and} \quad
  \bm{v}_2 = \Bigl[ \Re (\im\gamma'), \Im (\im \gamma'), \Re (\im \Omega_s)\Bigr],
\end{equation}
where $\Omega_s$ is the derivative of $\Omega$ with respect to the parameterization variable $s$. The real operator in the third component of both vectors can be changed if a different three-dimensional representation is desired. In this way, the surface mesh shown in Fig.~\ref{fig:bdcont_surface} is an accurate representation of the local topology of $\Omega$.

Notice that in contrast to Fig.~\ref{fig:reflection}, in Fig.~\ref{fig:bdcont_surface}, we prefer to plot $\Omega$ since it is readily available from the calculations. However, the $z$ values can be retrieved through \eqref{hodograph}. In this paper, we often switch between different solution measures and thus different visualizations of the Riemann surface. For example, we may prefer to plot the real or imaginary parts of the profile $z$, or the logarithmic hodograph variable, $\Omega = \log (1/z')$, or the complex velocity, $\exp(\Omega)$. Because the solution must always be of square-root type in $z$, then continuation properties between $z$ and $(1/z')$ will be analogous. Depending on whether the real or imaginary part is taken, visualizations of $\Omega$ may display a logarithmic branch point. Generally, our strategy is to choose a measure where the relevant properties (\emph{e.g.} branch cuts or singularities) are easily and unambiguously identified.

\section{Notation for the singularities and Riemann sheets}

\begin{figure} \centering
\includegraphics{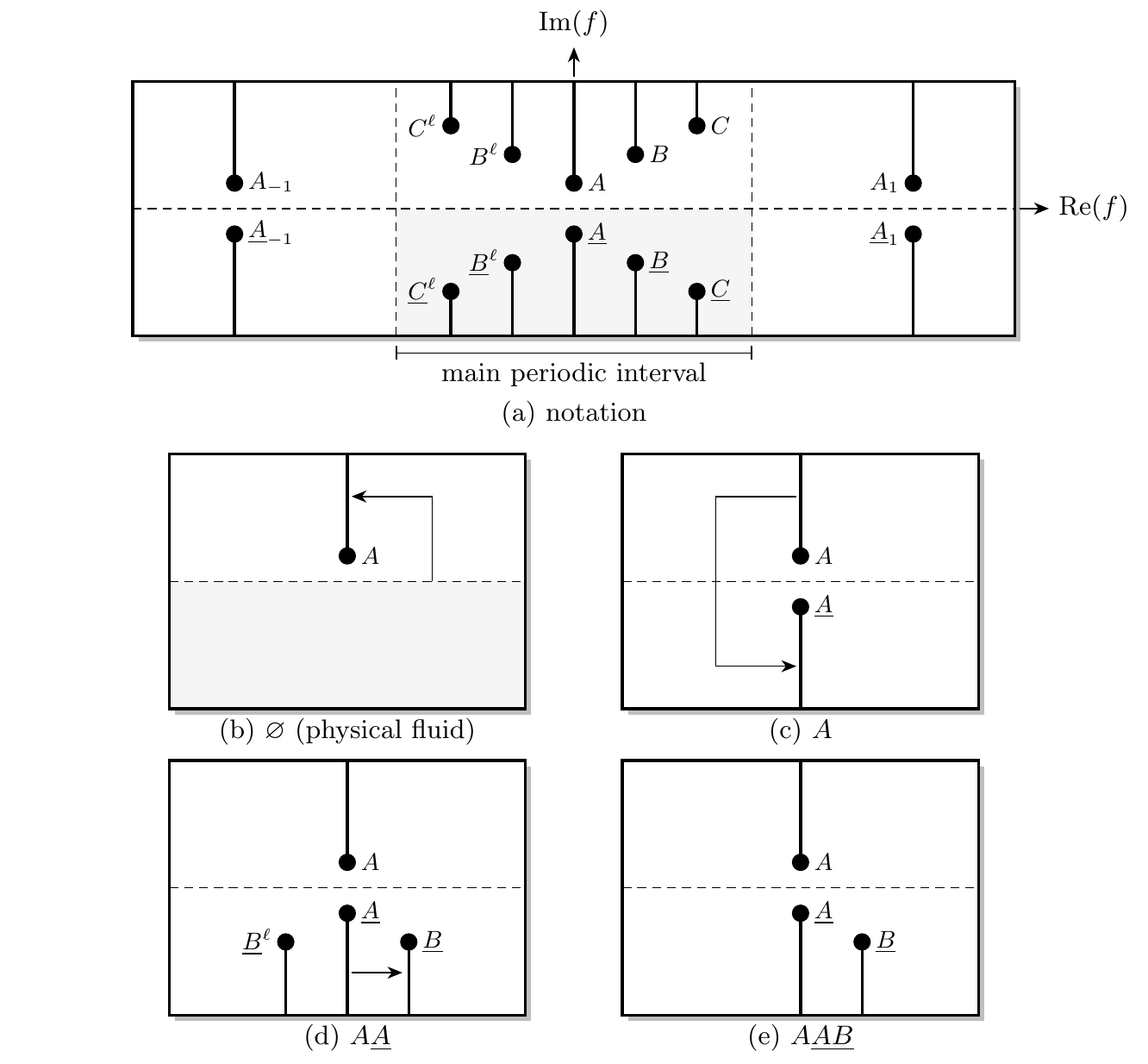}
\caption{(a) Notation for the singularities; the sequence illustrated in (b) through (d) shows the continuation paths (arrow) to reach the Riemann sheet A\uline{A}\uline{B} (e). Branch cuts are taken upwards/downwards for the upper/lower half-planes. \label{fig:notation}}
\end{figure}

In this section, we develop a notation to describe the singularity structure and collection of Riemann sheets for the complexified Stokes wave. For this, we must anticipate the results presented in \S\S\ref{sec:riemann}--\ref{sec:global}; there, we shall demonstrate that when viewed in the $f$-plane, a set of complex singularities of the Stokes wave will form two symmetrical V-shaped arrays, as illustrated in Fig.~\ref{fig:notation}(a). Singularities are initially marked with an upper-case alphabetical letter, \emph{e.g.} $\A$, $\B$, $\C$, and so forth. Those in the lower plane are distinguished by an underline, such as $\Au$, while those in the left plane are distinguished by a superscripted $\ell$, as in $\B^\ell$. 

In fact, the two main crest singularities, $\A$ and $\Au$, will also appear in a periodic fashion, and such copies will be marked by a subscripted index, such as $\A_{\pm 1}$, $\A_{\pm 2}$, and so forth, connected with the singularity $\A$ in the main interval. We have devoted a special section in \S\ref{sec:global} to explain the periodic properties of the singularities; however, our principal challenge in this paper will be first to understand the singularities within the main periodic interval. 

It is important to remark that singularities in the Stokes wave may not necessarily---and indeed they do not---exist on all Riemann sheets. The illustration in Fig.~\ref{fig:notation}(a) is effectively a projection of all such singularities onto the two-dimensional $f$-plane. Thus, depending on the path of continuation that is used to reach a given point in the $f$-plane, the singularity may or may not exist at that given location. 

In order to complete the description of the Riemann surface, we must thus describe how each individual Riemann sheet is reached using a path of analytic continuation. To this end, we introduce a sequence of letters for each individual sheet. The letters correspond to the singularities introduced above, with the additional symbol `$\varnothing$' to refer to the main Riemann sheet on which the physical fluid is found, and from which the analytic continuation begins. Consider, for example,
\begin{equation}
   \text{Riemann sheet $\A\Au\Bu$}.
 \end{equation} 
 Each of the three letters in the sequence corresponds to crossing a branch cut from the associated singularity in the positive (counterclockwise) sense. Thus, Riemann sheet $\A\Au\Bu$ is reached by beginning on $\varnothing$, and then encircling the singularities at $\A$, $\Au$, and $\Bu$, one after the other, as illustrated in Figs.~\ref{fig:notation}(b)--(e). As established in \S\ref{sec:singgen}, crossing a branch cut in the positive direction is equivalent to crossing in the negative direction; however, for precision of our numerical results, we will always consider the above notation as corresponding to a positive crossing.

\section{The crest singularity and its mirror image} \label{sec:riemann}

\noindent The purpose of this section is to apply the two methods for analytic continuation developed in \S\ref{sec:analcont} in order to develop a visualization of the Riemann surface for the Stokes wave, and to verify the presence of the proposed singularities. We will generally depend on the boundary integral scheme in \S\ref{sec:bdint}, which is unrestricted in the path of analytic continuation. However, the reflection scheme in \S\ref{sec:reflect} was used throughout to verify the accuracy of paths that were sufficiently simple (notably paths in \S\ref{sec:riemanncrest}). 

\subsection{The main Riemann sheet $\emptyset$ and the crest singularity $\A$} \label{sec:riemanncrest}

Previously during the explanation of the reflection scheme of \S\ref{sec:reflect}, we presented clear visual evidence of the main crest singularity, as it appears in the surface plot of Fig.~\ref{fig:reflection}. This singularity, which we refer to as `$\A$', lies on the main Riemann sheet $\varnothing$.

Let us confirm that $\A$ is indeed a square-root branch point. Using the boundary integral scheme, we analytically continue along a path that begins from the origin, tends upwards to $f_1 = 0.05\im$, and then performs two rotations around an approximated singularity location for $\A$ at $f = f_\A$. For the purpose of the visualization, we use a perturbed circle parameterized by 
\begin{equation} \label{eq:pertcirc}
\begin{split}
f(\theta) = |f_1 - f_{\mathrm{A}}| \Bigl[1 + \delta \sin\left\{ \tfrac{1}{2}(\theta_0 + \theta)\right\}\Bigr] \e^{\i (\theta_0 + \theta)}, 
\end{split}
\end{equation}
where $\theta_0 = \Arg(f_1 - f_{\mathrm{A}})$, and $0 \leq \theta \leq 4\pi$.

\begin{figure} \centering
\vspace*{5.0\baselineskip}
\includegraphics{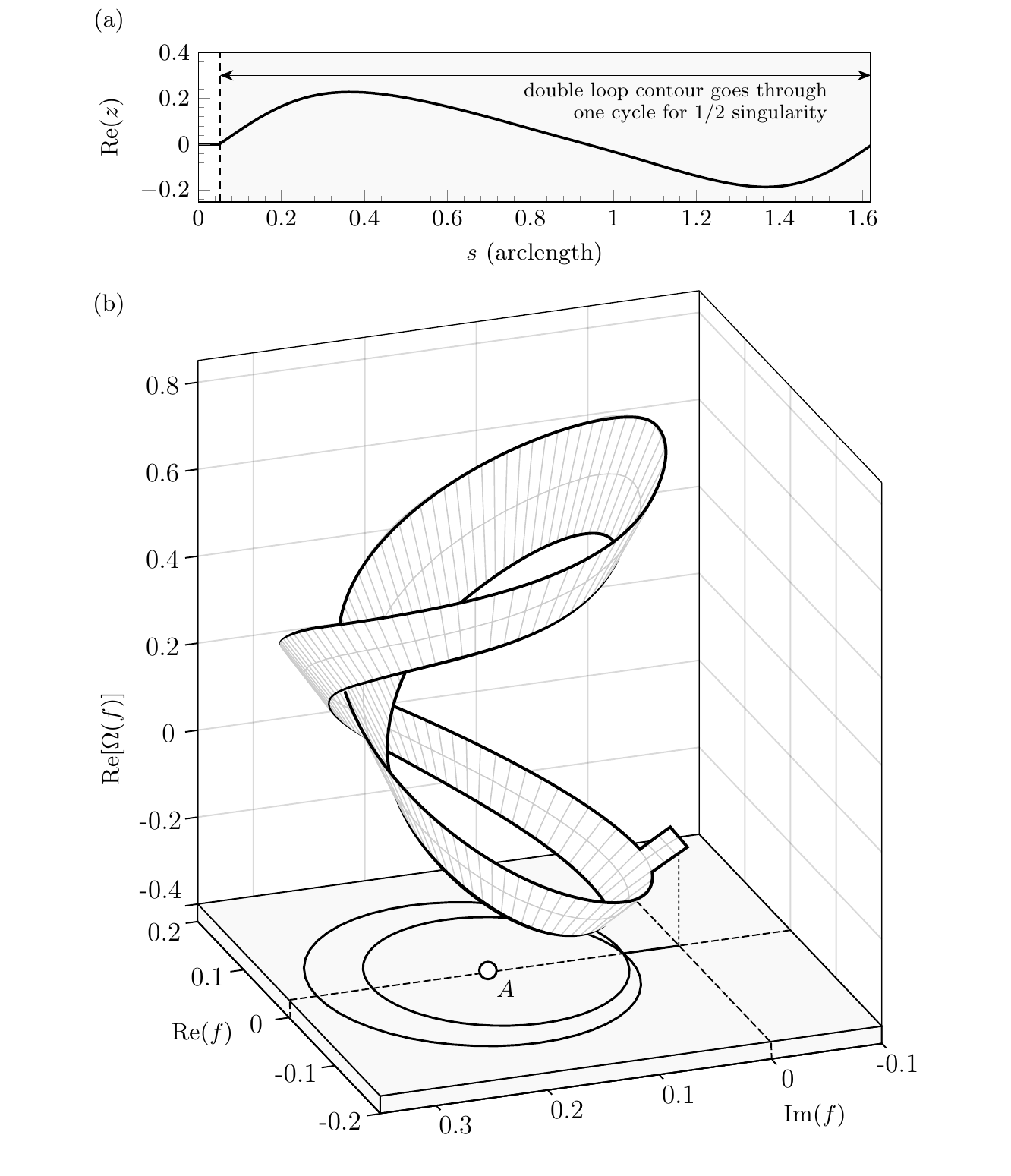}
\caption{(a) $\Re(z)$ as a function of the arclength, $s$ of the path of continuation; (b) the Riemann surface in $(\Re f, \Im f , \Re \Omega)$-space showing that two rotations around the \uline{A} singularity brings the path back to itself. The surface is computed using the boundary integral method and corresponds to $\ep = 0.3$ Stokes wave. The singularity is located at $f\approx 0.176\im$. \label{fig:surfA}}
\end{figure}

The numerical result, shown in Fig.~\ref{fig:surfA}, corresponds to the Stokes wave with steepness parameter $\ep = 0.3$. The three-dimensional representation in Fig.~\ref{fig:surfA}(b) shows the square-root behaviour as projected into $(\Re f, \Im f, \Re \Omega)$-space. In particular, note that the point, $f_1 = 0.05\im$, which is shared by the two circular orbits, takes two distinct values at the beginning of the first and second orbits, before finally returning to its original value at the end. This behaviour is to be expected for a path that has traversed both Riemann sheets near a square-root branch point. 

Once the numerical values along the path are known, we may then apply the integral equation \eqref{f0calc} to calculate a more accurate location for the singularity. For example, for $\ep = 0.3$, this computation yields $f_\mathrm{A} \approx 0.1756\im$. As shown in Fig.~\ref{fig:stokesprofile}(c), in the limit $\ep \to 0$, the singularity tends to $\im \infty$, while in the limit $\ep \to 1$, the singularity tends to the origin. The general dependence of the singularity location on $\ep$ will be presented later in \S\ref{sec:global}. 

Not only do these numerical results allow us to confirm the square-root nature of the solution, but we may also confirm the precise multiple of the square-root scaling. In \S\ref{sec:sing}, we argued (from \citealt{grant_1973_the_singularity}), that 
\begin{equation} \label{eq:zA}
  z \sim a_0 + a_1 (f - f_\mathrm{A})^{1/2},
\end{equation}
near the singularity. We now examine Fig.~\ref{fig:surfA}(a), which plots the real part of $z$ as a function of the arc-length along the path. As the path undergoes the two circular orbits, $\Re(z)$ undergoes one periodic cycle. This confirms the $\frac{1}{2}$ power in \eqref{eq:zA}, as opposed to some other $(f - f_\mathrm{A})^{m/2}$ power. This same scaling will not be true of the mirror singularity in the lower half-plane.

\subsection{The adjacent sheet and its mirror singularity $\protect\underline{A}$}

In \S\ref{sec:singcrest}, we explained why a singularity might be expected at the mirror point $-f_{\A}$ on the Riemann sheet adjacent to $\varnothing$. The conjecture, which was first proposed by \cite{tanveer_1991}, is based on the reflection property of \eqref{eq:bernoulli_reflected_hidden} and the non-analyticity of $\PP$ and $\QQ$ due to $f_{\A}$. Let us call this mirror singularity `$\Au$', with the underline to remind us of its placement in the lower half-plane. 

\begin{figure} \centering
\includegraphics[width=1.0\textwidth]{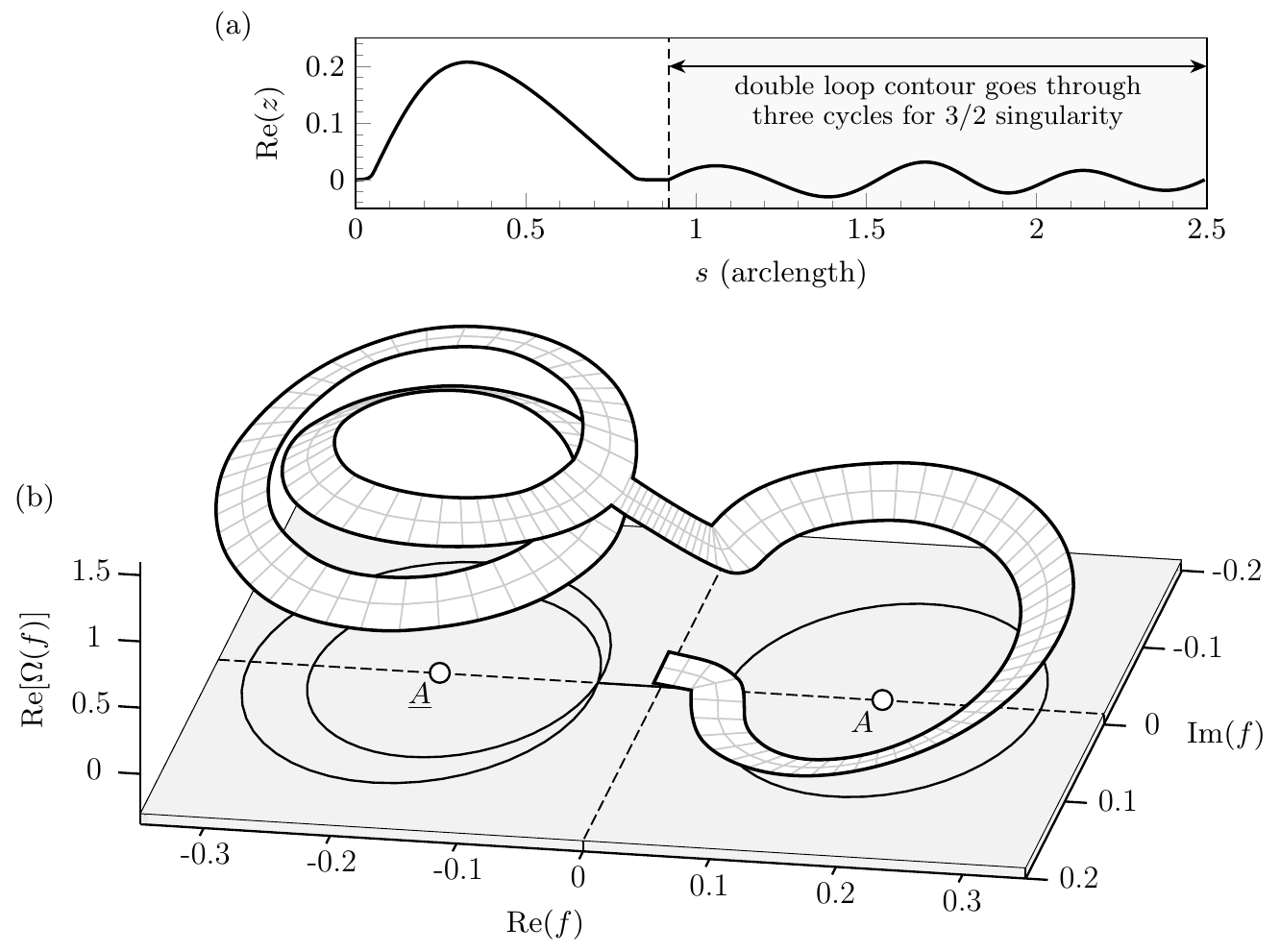}
\caption{(a) $\Re(z)$ as a function of the arclength, $s$ of the path of continuation. (b) The Riemann surface in $(\Re f, \Im f , \Re \Omega)$-space showing that two rotations around the \uline{A} singularity brings the path back to itself. The surface is computed using the boundary integral method and corresponds to the $\ep = 0.3$ Stokes wave. The two singularities are located at $f\approx \pm 0.176\im$. \label{fig:surfAA_}}
\end{figure}

Consider the case of $\ep = 0.3$. Using the boundary integral scheme, we perform the following two motions: first, beginning from $f = 0$ on the main sheet $\varnothing$, we perform a positive (counterclockwise) orbit around $\A$, and then return to the origin. This brings us to the adjacent Riemann sheet. Next, we analytically continue downwards to $f_1 = -0.05\im$, and then perform two positive rotations centred on $-f_\A$ using the analogous parameterization to \eqref{eq:pertcirc}. The result is shown in Fig.~\ref{fig:surfAA_}(b) and confirms that two rotations brings the function back to its original value. Calculation of the integral \eqref{f0calc} confirms that the singularity is indeed located at the reflected point.

In \S\ref{sec:singcrest}, we demonstrated that if such the mirror singularity was found, then the solution must of the form
\begin{equation}
  z \sim b_0 + b_1 (f + f_\mathrm{A})^{3/2},
\end{equation}
as opposed to the $\frac{1}{2}$ scaling of \eqref{eq:zA}. This is confirmed with Fig.~\ref{fig:surfAA_}(a) where $\Re[z(f)]$ is observed to undergo three nearly-periodic cycles instead of the single cycle in Fig.~\ref{fig:surfA}(a).

One might ask: \emph{what happens if the path does not encircle a singularity?} For this, a comparison can be drawn to the analytic continuation shown in Fig.~\ref{fig:reflection} where the rectangular path (shown solid) does not encircle the $\A$ point, and therefore the continuation into the lower half-plane returns to the physical fluid (where it is analytic everywhere and single-valued). Had there been no singularity in the lower half plane, we would expect all paths of continuation to be single-valued along closed orbits. 

\section{The existence of diagonal singularities} \label{sec:diag}

By now, we have confirmed the existence of the two main singularities. Now we will show that further singularities emerge when proceeding deeper into the surface. These new singularities are accessed through paths that traverse branch cuts from $\A$ and $\Au$, and are positioned diagonally from $\Au$---more specifically, they tend to a constant argument as $\ep \to 0$. As explained in \S\ref{sec:singgen}, the initial emergence of these diagonal singularities is somewhat out of \emph{chance}; their existence depends on initial conditions of the Stokes wave and they are associated with locations where $\PP = 0$. However, once they appear, they will also cause the appearance of further mirror singularities, where $\PP \neq 0$. 

In fact, after submission of this work, we became aware that within a preprint, \cite{lushnikov2015branch} had independently conjectured the existence of such diagonal singularities. The conjecture was based on an observation that the sudden transition of powers (\emph{cf.} p.~\pageref{grantquote}) from two singularities on the imaginary axis to the single highest-wave singularity could be achieved using an infinite sequence of nested square roots; the resultant function then admits off-axis singularities. 

\subsection{Diagonal singularities $\Bu$ and $\Bu^\ell$}

As an example, consider the case of $\ep = 0.5$ and the Riemann sheet $\A\Au$. In order to reach this sheet, we have taken the continuation path shown in the lower two-dimensional plane of Fig.~\ref{fig:surfB}(b); it begins on the positive $\Re(f)$ axis and forms an elliptical orbit around $\A$ and $\Au$. We have chosen to illustrate the Riemann surface using its projection into $(\Re f, \Im f , \Re 1/z')$-space, as this provides the most clear visualization. Unlike the previous visualizations, where we have only meshed along the continuation path, here, we provide a more complete surface mesh for the Riemann sheet. The surface mesh is created by first encircling $\A$ and $\Au$, and then returning to the real $f$-axis. From the axis, the continuation moves in straight vertical lines (similar to the scheme to create Fig.~\ref{fig:reflection}). This process has the effect of imposing vertical branch cuts from the singularities. 

On the Riemann sheet shown in Fig.~\ref{fig:surfB}(b), we observe the usual $\A$ singularities and its periodic copies. Only $\A_{-1}$ and $\A_1$ are seen, but other copies are shifted a unit length away from one another. Notice that periodic copies of $\Au$ do not appear; we shall explain the periodicity structure in more detail in \S\ref{sec:global}. Most importantly, observe the presence of the two branch cuts clearly visible in the lower half-plane. These branch cuts emerge from two points located symmetrically about the imaginary axis:
\begin{equation} \label{fB}
  f_{\Bu} \approx 0.557 - 0.396\im \quad \text{and} \quad
  f_{\Bu^\ell} \approx -0.557 - 0.396\im.
\end{equation}
The doubly-rotated path in Fig.~\ref{fig:surfB}(b) confirms the square-root nature of the new diagonal singularity, $\Bu$, and examination of Fig.~\ref{fig:surfB}(a) verifies that the singularity is of $\frac{1}{2}$ type, as anticipated by \eqref{Psing}. 

A plot of the locations of the $\B$-type singularities is shown in Fig.~\ref{fig:Blocations}(a), along with the dependence of the singularity magnitudes and arguments as a function of $\ep$, seen in subplots (b) and (c). It is confirmed that as $\ep \to 0$, $|f_\B| \to \infty$, while $\Arg f_\B$ tends to a constant. The numerical computations are restricted to $\ep \leq 0.6$, since above this value it becomes difficult to preserve accuracy of the singular integral. Based on Fig.~\ref{fig:Blocations}, however, we would expect that $|f_\B| \to 0$ as $\ep \to 1$, but the limiting angle is unclear.

\subsection{Mirror diagonal singularities $\B$ and $\B^\ell$}

We may confirm that as $\Bu$ and $\Bu^\ell$ are approached, $\PP \to 0$, which follows our argument leading to \eqref{Psing}. Thus, from the same argument, it is expected that the singularities \eqref{fB} will cause the emergence of mirror singularities (reflected about the origin) due to the non-analyticity of $\PP$ and $\QQ$. In fact, these mirror singularities can be more simply argued based on the reflection scheme of \S\ref{sec:reflect}. 

\afterpage{
\clearpage
\begin{landscape}
\thispagestyle{lscape}
\pagestyle{lscape}
\begin{figure} \centering
\includegraphics[angle=90, scale=0.9]{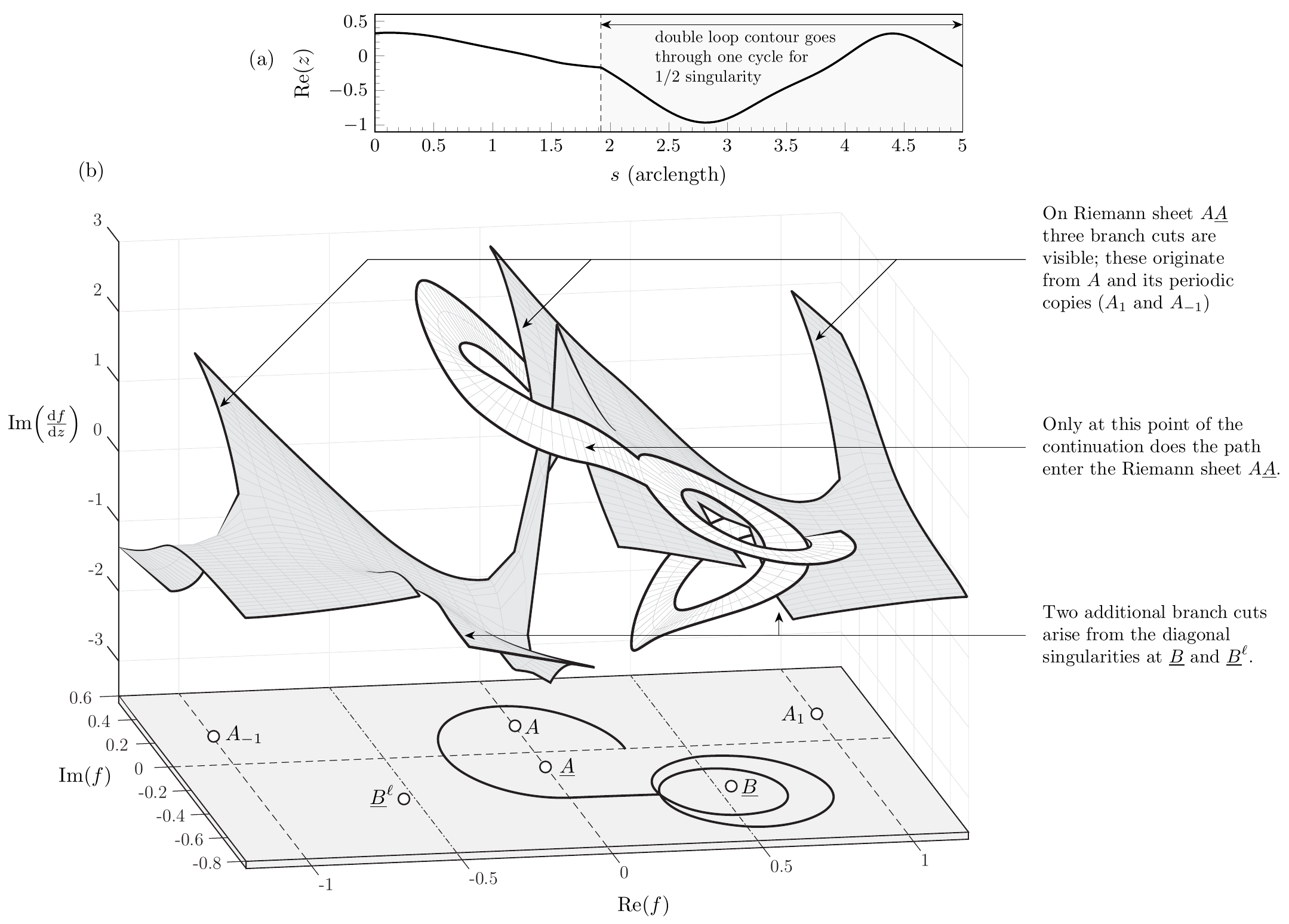}
\caption{These two pictures confirm the existence of the diagonal singularities $
\{\Bu, \Bu^\ell\}$ for the Stokes wave with $\ep = 0.5$. Continuation follows a path that encircles $\A$ and $\Au$, and then the purported singularity at $\Bu$. (a) The profile of $\Re(z)$ along the path confirms the singularity scaling. (b) The surface mesh for $\A\Au$ and continuation mesh showing the double rotation around $\Bu$. 
\label{fig:surfB}}
\end{figure}
\end{landscape}
}

Consider a rectangular path, similar to Fig.~\ref{fig:reflection}, that orbits both $A$ and $\Au$. In the first quadrant (top-right) of sheet $\A\Au\A$, $z(f)$ is derived based on values of $z(-f)$ from the third quadrant (bottom-left). Consequently, $z(-f)$ lies on Riemann sheet $\A\Au$, which contains the $\Bu^\ell$ singularity with $\PP \to 0$. Thus, by \eqref{eq:bernoulli_reflected}, it follows that there is a singularity at the reflected point in the first quadrant. These mirror singularities, $\B$ and $\B^\ell$ are confirmed in the later Fig.~\ref{fig:picto}(f), and we may also verify that they are of the $\frac{3}{2}$ type in \eqref{Psing}.

\subsection{Further diagonal singularities}

The diagonal structure does not end there. In fact, it can be verified that there are two further singularities of $\C$-type: $\{\C, \C^\ell, \Cu, \Cu^\ell\}$. For example, if we perform two elliptical orbits instead of the single one of Fig.~\ref{fig:surfB}(b), we would arrive at Riemann sheet $\A\Au\A\Au$. On this sheet, we discover a similar topology to the one for $\A\Au$, but with the diagonal set $\{\Bu, \Bu^\ell, \Cu, \Cu^\ell\}$. These $\C$-type singularities are diagonally-shifted further than for the previous case of the $\B$-type singularities, and they too will induce the mirror set $\{\C, \C^\ell\}$. 

If we strictly confine ourselves to climbing the $\A\Au$ structure, then the pattern of singularities is clear: each additional tour adds a further set of diagonal points. Thus for example, the three-tour Riemann sheet $\A\Au\A\Au\A\Au$ contains $\{\Bu, \Bu^\ell, \Cu, \Cu^\ell, \Du, \Du^\ell\}$. The more difficult challenge is to characterize the surface once we proceed onto Riemann sheets accessed through the diagonal points. This will chiefly be our task in \S\ref{sec:global}.

\section{The global structure of the Riemann surface} \label{sec:global}

\noindent In this section, our goal is to provide a basic list of rules and observations that characterize the global structure of the Riemann surface for the Stokes wave. As it turns out, a full description of the surface will prove to be quite difficult, but we hope to give the reader a taste of the daunting complexity of the singularity structure.

Earlier in Fig.~\ref{fig:notation}(a), we had presented our notation describing the main crest singularity $\A$, its mirror image $\Au$, and also the main diagonal singularities that form the two V-shaped arrangements, $\V = \{\B, \B^\ell, \Bu, \Bu^\ell, \C, \C^\ell, \Cu, \Cu^\ell, \ldots\}$. Our numerical results have indicated that beyond these `first diagonals', further singularities (diagonals of diagonals) arise once the path of continuation encircles the former set. Newer diagonals do not share the same symmetry properties as $\V$.

\subsection{Symmetry-breaking paths of continuation} \label{sec:symmetry}

Our discussion will centre on numerical results for the Stokes wave with $\ep = 0.5$. In Fig.~\ref{fig:picto}(a), we plot the locations of the singularities discussed earlier in \S\ref{sec:riemann} and \S\ref{sec:diag} using a white node. Contour plots for $\Re \Omega = \Re (\tau - \im \theta)$ are shown in subfigures (b)--(j). We have chosen this solution measure as it is one for which the singularities are apparent from the shape of the contours. The white nodes in the smaller subfigures correspond exactly to those in Fig.~\ref{fig:picto}(a), and the black nodes mark additional singularities that arise once symmetry is broken in the path of analytic continuation. 

This notion of symmetry breaking is a central reason for the difficulty in establishing the full Riemann surface. Earlier we presented a method of analytic continuation that was based on formula \eqref{eq:bernoulli_reflected}. Here, the reflective property of the equation indicates that if there exists a singularity at $f = f^*$, then there is likely a singularity as well at $-f^*$. However, this criterion only applies if, by the time the continuation path reaches $-f^*$, the former $f^*$ is still singular \emph{on the same Riemann sheet}. This is certainly true for the transition between Riemann sheets $\varnothing$ and $\A$, where singularity $\A$ is located on the imaginary axis on both sheets, and hence induces the creation of the mirror singularity $\Au$. However, in general, we must expect that paths of continuation that encircle a diagonal singularity to break the symmetry of the Riemann sheet; thus, arguments that depend on reflection will no longer directly apply. 

\begin{figure} \centering
\includegraphics{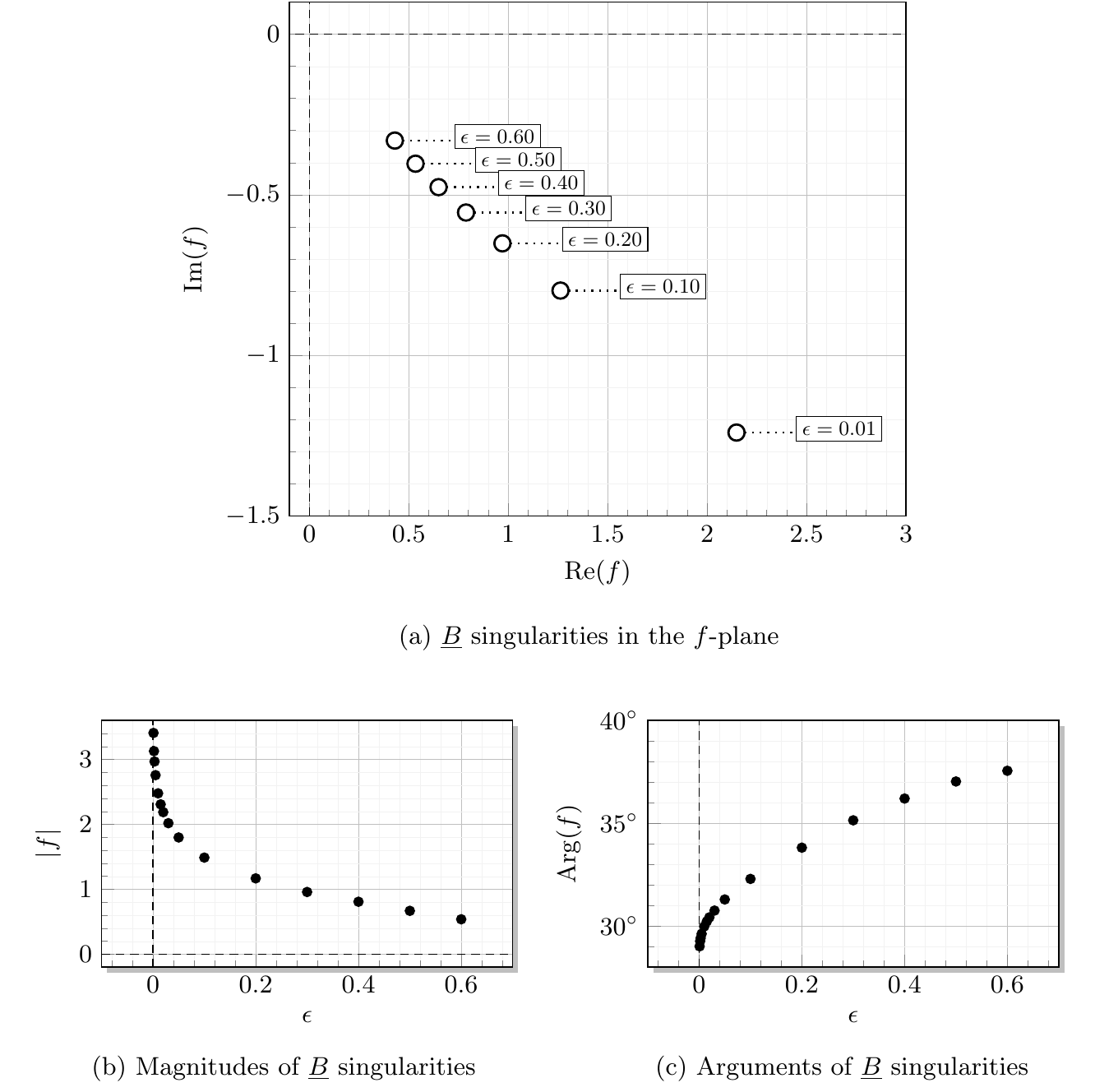}
\caption{Location of the singularities $\{\B, \B^\ell, \Bu^\ell, \Bu\}$ (from first to fourth quadrants) calculated using the integral \eqref{f0calc}. Singularities $\Bu$ and $\Bu^\ell$ are found on Riemann sheet $\A\Au$. Singularity $\B^\ell$ is found on $\A\Au\Bu$ and singularity $\B$ is found on $\A\Au\Bu^\ell$. \label{fig:Blocations}}
\end{figure}

\subsection{Rules and observations}

By now, the first two observations are well known:
\begin{quotation}
\emph{(1) On the main sheet $\varnothing$, there is only the single crest singularity $\A$ and its periodic copies, $\A_{\pm i}$, $i = 1, 2, 3, \ldots$}

\emph{(2) Two rotations about any singularity returns the path to its original value; clockwise and counterclockwise rotations are equivalent.}
\end{quotation}

\noindent The first observation is verified in Fig.~\ref{fig:picto}(b), and in numerous points throughout this work. The second observation follows from the fact that, as discussed in \S\ref{sec:sing}, the only possible dominant balance in \eqref{eq:bernoulli_reflected_hidden} must necessarily yield singularities of square-root-type. We have also made mention of the periodic copies of $\A$, of which $\A_1$ and $\A_{-1}$ can be seen in the figure. This leads us to:

\begin{figure} \centering
\includegraphics{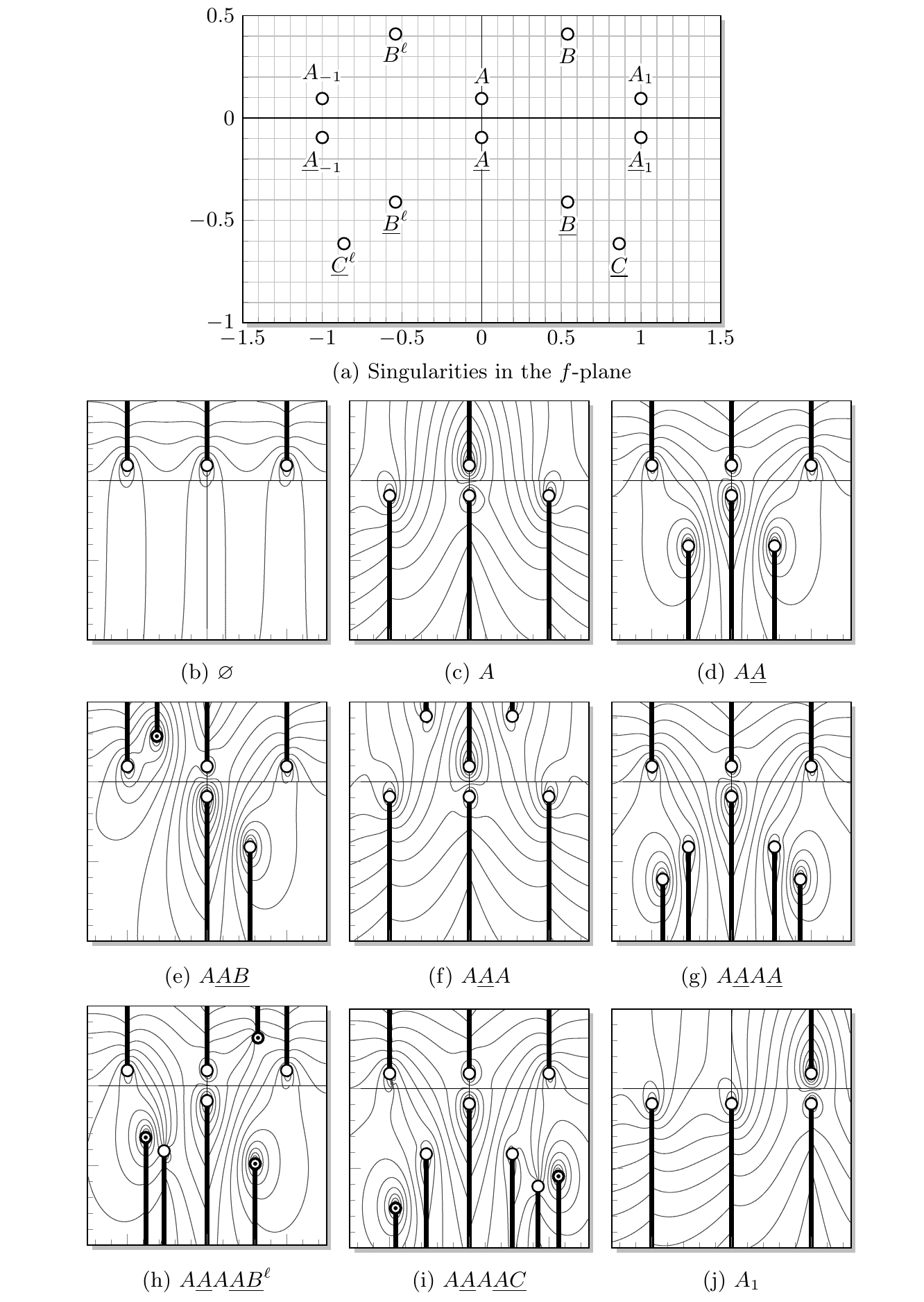}
\caption{Riemann sheets for the $\ep = 0.5$ Stokes wave. The white nodes mark the crest singularity and its two periodic copies, the mirror singularity, and the first diagonal singularities. Black nodes mark singularities arising from diagonals. The contours correspond to $\Re \Omega = \Re(\tau - \im \theta)$. \label{fig:picto}}
\end{figure}

\begin{quotation}
\emph{(3) Periodic copies of $\A$ and $\Au$ can appear on certain Riemann sheets; when this occurs, they appear at the points $\pm f_\A + \mathbb{Z}$.}
\end{quotation}

\noindent Note that while the physical water wave is required to be periodic, there is no such limitation on the analytic continuation. However it is a consequence of the two methods of continuation explained in \S\ref{sec:analcont} that certain periodic properties may still be preserved. As shown in Fig.~\ref{fig:picto}(b), on $\varnothing$, there will be periodic copies of the main singularity, $\A$, separated by unit lengths. In fact, this can be more easily seen by using the circular map given by $t = \exp(-2\pi\i f)$. This maps the lower half-$f$-plane to the interior of a unit circle in the $t$-plane, and all singularities $f_\A + \mathbb{Z}$ to a single point $t_\A$. By similar logic, the mirror singularity $f_{\Au}$ will also be accompanied by periodic copies on certain Riemann sheets, as in Fig.~\ref{fig:picto}(c). 

\begin{quotation}
\emph{(4) By solely encircling the combination $\A\Au$, diagonal singularities will appear in pairs, from $\{\Bu, \Bu^\ell, \Cu, \Cu^\ell, \Du, \Du^\ell, \ldots\}$, with a new type for each orbit of $\A\Au$. From the chain of $\A\Au$ orbits, encircling $\A$ once further will cause the upper diagonal singularities $\{\B, \B^\ell, \C, \C^\ell, \D, \D^\ell, \ldots\}$ to appear in an analogous fashion.}
\end{quotation}

\noindent Thus for example, on sheet $\A\Au$, singularities $\{\Bu, \Bu^\ell\}$ appear. On sheet $\A\Au\A\Au$, singularities $\{\Bu, \Bu^\ell, \Cu, \Cu^\ell\}$ appear, and so forth. This statement was the focus of \S\ref{sec:diag},and we also see the behaviour in Figs.~\ref{fig:picto}(d) and (g). Adding the additional circling of $\A$ produces the mirror singularities; thus $\A\Au\A$ has singularities $\{\B, \B^\ell\}$, as shown in Fig.~\ref{fig:picto}(f).

Indeed there are a plethora of other rules that may derive with the help of the continuation schemes discussed in \S\ref{sec:analcont}. However, we will end here with one final observation---the most vague and open of the ones we have proposed. 
\begin{quotation}
\emph{(5) Circling a diagonal singularity causes the generation of further singularities on the new sheet. The locations of these singularities are unpredictable and not subject to symmetry arguments in the $f$-plane.}
\end{quotation}

\noindent The observation was discussed in \S\ref{sec:symmetry}, and can be seen in Fig.~\ref{fig:picto}(e, g, h). In such figures we see that upon circling a diagonal singularity, new singularities may arise (marked as a dark node). Such singularities arise in a non-symmetric fashion.

\section{Conclusion}

Using numerical computations of the analytic continuation of the Stokes wave, we have confirmed the existence of the main crest singularity, as shown by \cite{grant_1973_the_singularity}, \cite{schwartz_1974}, and many others. We have also confirmed the existence of its mirror image, as first conjectured by \cite{tanveer_1991}. Further exploring deeper layers of the Riemann surface reveals a countably infinite number of diagonal singularities. 

\section{Discussion} \label{sec:discussion}

Throughout his numerous correspondences with William Thompson (Lord Kelvin), Stokes made frequent note of his hesitation that the highest possible wave would indeed exhibit a crest angle of $120^\circ$. However by September 1880, he seemed to have convinced himself of the theory's veracity: 
\begin{quotation}
\noindent \emph{You ask if I have done anything more about the greatest possible wave. I cannot say that I have, at least anything to mention mathematically. For it is not a very mathematical process taking off my shoes and stockings, tucking up my trousers as high as I co[u]ld, and wading out into the sea to get in line with the crests of some small waves that were breaking on a sandy beach [\ldots] it did seem to me that the waves began to break while their sides still made only a blunt angle, a good deal less than $90^\circ$. I feel pretty well satisfied that the limiting form is one presenting an edge of $120^\circ$. \citep[p.498]{stokes_sept1880}}
\end{quotation}

While Stokes may have been able to reassure himself through direct observations of physical waves, for us, such interactions are more difficult---not least because the singularities we study are confined to the complex plane. However, a connection with the physicality of the problem need not be out of reach: through our construction and exploration of the Riemann surfaces, we have sought to offer the same sense of discovery that Stokes had found that day, wading into the sea.

It is furthermore remarkable that, perhaps the simplest possible problem involving the study of a nonlinear water wave, a topic of study that dates back over a century and a half, continues to confound us in the present day with new challenges. While we have attempted to highlight the complexity of the underlying issues, there are naturally three questions that follow from this work. 

\emph{(1) What is the complete mathematical description of the Riemann surface for the Stokes wave?} An exhaustive search of the individual Riemann sheets of the surface is yet to be undertaken. Here, we have sought to describe the patterns and connections between the branch-cut transitions for the first few Riemann sheets adjacent to the physical fluid. While this has led to the particular example shown in Fig.~\ref{fig:picto}, and our proposed rules and observations of \S\ref{sec:global}, a more complete description of the full surface topology will be the subject of forthcoming work.

\emph{(2) What are the asymptotic properties of the singularities in the two limits of $\ep \to 0$ (small waves) and $\ep \to 1$ (steep waves)?} Though the case $\ep \to 0$, may seem relatively benign, we still do not understand the various properties of the singularities in this limit. For example, it seems from Fig.~\ref{fig:Blocations} that the diagonal singularities tend to infinity along a constant angle as $\ep \to 0$. Can these angles, as well as other properties of the Riemann surface be analytically predicted? There are strong connections between these observations and the occurrence of periodic arrays of singularities in other studies on nonlinear differential equations (see \emph{e.g.} \citealt{chapman_2013_exponential_asymptotics, costin_2001}).

Clearly, in the limit $\ep \to 1$, the cancellation of the leading square-root power cannot be described using something as trivial as 
\begin{equation} \label{wrongz}
  z \sim \Bigl[ a_0 + a_1 (f - f_\A)^{1/2} + \ldots \Bigr] + 
  \Bigl[ \tilde{b}_0 - a_1 (f + f_\A)^{1/2} + \ldots \Bigr],
\end{equation}
between the main crest singularity and its mirror image, and with $f_\A \to 0$. For one reason, we have established that the mirror singularity is of order $\frac{3}{2}$. For another, encircling both singularities and climbing the $\A\Au$ structure does not return the continuation to its original value---the topology of the Riemann surface is infinitely sheeted along this path. In contrast, two rotations in \eqref{wrongz} will return the solution to its original value. Thus, any asymptotic description of the matching region between outer (away from the singular crest) and inner solutions cannot be composed of any finite number of square-root sums, as in \eqref{wrongz}. Indeed this was the observation from a preprint by \cite{lushnikov2015branch}, who conjectured that an infinite product of increasingly nested square roots might explain the recombination of the two main crest singularities to Stokes' single crest singularity. As noted by the author, such a functional form would necessarily have additional off-axis singularities. It is intriguing that we have found such singularities here as well. 

The work by \cite{longuet-higgins_1977_theory_of,longuet-higgins_1978_theory_of} studying the `almost-highest wave' is most relevant, and we hope to undertake a comprehensive study reviewing the connections between the previous work and our present paper. Problems with coalescing singularities often reveal challenging asymptotic structures for study \citep{chapman_2013_exponential_asymptotics, trinh_2015_exponential_asymptotics}.

\emph{(3) What are the physical and practical consequences of the newfound singularity structure?} Finally, the question most relevant to the practitioner (numerical, theoretical, and experimental) is what we may take away from this study, as it relates to real-life consequences of Stokes wave theory. We had already discussed, in \S\ref{sec:singimportance}, many of the issues where better understanding of the singularity structure would illuminate. As many theories on time-dependent instabilities rely upon the interaction of the nearest singularity to the free surface, the question of how to account for the coalescence of the (presumed) infinity of singularities as $\ep \to 1$ seems to be of paramount importance. In addition, we also highlight intriguing questions on the role of Stokes wave singularities in the development of capillary ripples riding on steep gravity waves [\emph{cf.} \cite{longuet_1995}]. As shown in \cite{trinh_2013_new_gravity-capillary,trinh_2013a_new_gravity-capillary}, it is expected that the singular limit of small surface tension will present a challenging route for further study.

\begin{acknowledgements}
\textbf{Acknowledgements}: We are indebted to Profs. S. Jonathan Chapman and Henry Woudhuysen (Oxford) for their insightful suggestions and for providing close support during the preparation of this work. PHT thanks Miss Christina Goodwin for providing editorial input. Both authors express gratitude to Lincoln College for financial support.
\end{acknowledgements}


\providecommand{\noopsort}[1]{}

\end{document}